\documentclass[11pt, a4paper]{article}
\usepackage{jheparxiv}
\usepackage[utf8]{inputenc}
\usepackage{amsmath}
\usepackage{amsfonts}
\usepackage{amssymb}
\usepackage{latexsym}
\usepackage{mathrsfs}
\usepackage{braket}		
\usepackage{graphicx}
\usepackage{color}
\usepackage{xcolor}
\usepackage{slashed}
\usepackage{twistor}
\usepackage[all]{xy}

\newcommand{\sa}{\mathsf{a}}

\newcommand{\vepsilon}{\varepsilon}

\newcommand{\sT}{\mathsf{T}}
\newcommand{\sA}{\mathsf{A}}
\newcommand{\sK}{\mathsf{K}}
\newcommand{\sG}{\mathsf{G}}
\newcommand{\sH}{\mathsf{H}}

\renewcommand{\d}{\mathrm{d}}

\subheader{\hfill \begin{tabular}{r} \texttt{IMPERIAL-TP-TA-2017-03}
 \\ \texttt{NSF-ITP-17-100} \end{tabular}}

\title{Amplitudes on plane waves from ambitwistor strings}

\author[a,c]{Tim Adamo,}
\author[b,c]{Eduardo Casali,}
\author[b,c]{Lionel Mason}
\author[b]{\& Stefan Nekovar}

\affiliation[a]{Theoretical Physics Group, Blackett Laboratory \\
        Imperial College London, SW7 2AZ, United Kingdom}

\affiliation[b]{The Mathematical Institute \\
        University of Oxford, Woodstock Road, OX2 6GG, United Kingdom}

\affiliation[c]{Kavli Institute for Theoretical Physics \\
        University of California, Santa Barbara, CA 93106, USA}

\emailAdd{t.adamo@imperial.ac.uk}
\emailAdd{[lmason,casali,nekovar]@maths.ox.ac.uk}

\abstract{In marked contrast to conventional string theory, ambitwistor strings remain solvable worldsheet theories when coupled to curved background fields. We use this fact to consider the quantization of ambitwistor strings on plane wave metric and plane wave gauge field backgrounds. In each case, the worldsheet model is anomaly free as a consequence of the background satisfying the field equations. We derive vertex operators (in both fixed and descended picture numbers) for gravitons and gluons on these backgrounds from the worldsheet CFT, and study the 3-point functions of these vertex operators on the Riemann sphere. These worldsheet correlation functions reproduce the known results for 3-point scattering amplitudes of gravitons and gluons in gravitational and gauge theoretic plane wave backgrounds, respectively.}

\begin{document}
\maketitle

\section{Introduction}

Using perturbative string theory to study physics in the presence of curved background fields is a highly non-trivial task. When coupled to generic curved background fields, the string worldsheet action becomes a complicated interacting 2d CFT which can only be studied perturbatively (e.g., \cite{AlvarezGaume:1981hn,Braaten:1985is}). Quantum consistency of the worldsheet theory imposes an infinite tower of higher-derivative constraints on the background fields, which at lowest order are the two-derivative equations of motion of field theory~\cite{Fradkin:1985ys,Callan:1985ia,Banks:1986fu,Abouelsaood:1986gd}. It is therefore exceptionally difficult to tell if a given background field configuration satisfies the full string equations of motion, or to compute interesting target space quantities, such as scattering amplitudes, in the resulting worldsheet CFT.

Over the years, some notable exceptions to the first of these difficulties have been found. Vacuum plane wave metrics were argued to be admissible NS-NS backgrounds for string theory due to the vanishing of their higher curvature invariants~\cite{Amati:1988sa,Horowitz:1989bv}. Supergravity solutions based on AdS (times a compact space)~\cite{Freund:1980xh,Schwarz:1983qr} or pp-waves~\cite{Blau:2001ne,Blau:2002dy} supported by Ramond-Ramond flux were argued to be admissible backgrounds for type II string theory on the basis of uniqueness and symmetry constraints for the integrable sigma models with these target spaces~\cite{Metsaev:1998it,Metsaev:2001bj,Metsaev:2002re,Bena:2003wd,Berkovits:2004xu,Arutyunov:2008if,Stefanski:2008ik}. These examples play a central role in the concept of holography~\cite{Maldacena:1997re,Gubser:1998bc,Witten:1998qj} and its plane wave limit~\cite{Berenstein:2002jq}. The class of supersymmetric sigma models with curved target spaces can be expanded to include various integrable deformations (c.f., \cite{Delduc:2013qra}), although it is not entirely clear if these deformations define consistent string theories~\cite{Arutyunov:2015mqj,Wulff:2016tju}.

Yet even with consistent string theories on curved backgrounds, writing explicit vertex operators or calculating worldsheet correlation functions has proved virtually impossible.\footnote{A notable special case where progress \emph{has} been made is for AdS$_3$, where the worldsheet theory is a $\SL(2,\R)$ WZW model~\cite{Maldacena:2000hw,Maldacena:2000kv,Maldacena:2001km}.} This is because the worldsheet model -- even if it is integrable -- remains an interacting CFT (as is the case for supersymmetric AdS backgrounds or vacuum plane waves), or because the worldsheet model is known only in Green-Schwarz form (as for the solvable pp-wave sigma models). Although some progress towards writing vertex operators on certain backgrounds has been made (c.f., \cite{Jofre:1993hd,Dolan:1999dc,Berkovits:2000yr,Chandia:2013kja}), there is still no intrinsically stringy computation of the 3-point function in a curved background.\footnote{It should be noted that worldsheet methods \emph{have} been used to compute correlators in certain limits~\cite{Minahan:2012fh,Bargheer:2013faa,Minahan:2014usa} or with special configurations of external states~\cite{Berkovits:2012ps,Azevedo:2014rva} in AdS backgrounds. Cubic string field theory has been used to study interactions on pp-wave backgrounds~\cite{Constable:2002hw,Spradlin:2002ar}.} This seems particularly remarkable given how much attention is paid to such backgrounds in the context of holography, where bulk observables are usually computed from field theory Witten diagrams rather than the string worldsheet.

\medskip

Our goal is to provide the first worldsheet calculation of vertex operators and 3-point functions on curved backgrounds using alternatives to standard string theory. Recently, it has been shown that certain chiral, constrained worldsheet theories -- known as \emph{ambitwistor strings} -- can be used to study perturbative field theory directly~\cite{Mason:2013sva}. In flat space, the genus zero worldsheet correlation functions of ambitwistor strings reproduce the full tree-level S-matrix for a wide array of field theories~\cite{Ohmori:2015sha,Casali:2015vta,Azevedo:2017lkz} in the scattering equations form of Cachazo-He-Yuan (CHY)~\cite{Cachazo:2013iea,Cachazo:2014xea}. These models lead to new representations of the S-matrix at higher loops in terms of worldsheet correlation functions at higher genus~\cite{Adamo:2013tsa,Casali:2014hfa,Adamo:2015hoa} or on the nodal Riemann sphere~\cite{Geyer:2015bja,Geyer:2015jch,Geyer:2016wjx}. Ambitwistor strings can be viewed as an alternative quantization of the null string~\cite{Casali:2016atr,Casali:2017zkz}, and have been studied in various different ways. This includes specializations to 4-dimensions~\cite{Geyer:2014fka,Bandos:2014lja}, applications to asymptotic symmetries and soft theorems~\cite{Adamo:2014yya,Geyer:2014lca,Lipstein:2015rxa,Adamo:2015fwa}, pure spinor generalizations~\cite{Berkovits:2013xba,Chandia:2015sfa,Jusinskas:2016qjd}, ambitwistor string field theory~\cite{Reid-Edwards:2015stz,Reid-Edwards:2017goq}, and even the study of space-time conformal invariance~\cite{Adamo:2017zkm}.

In the context of curved background fields, one particular feature of ambitwistor strings stands out: they remain free worldsheet CFTs even when coupled to generic background fields! This was proven in detail for the type II ambitwistor string (which describes type II supergravity on space-time), and quantum consistency was shown to follow from the supergravity field equations alone~\cite{Adamo:2014wea}. In other words, ambitwistor strings provide a description of non-linear field theory in terms of a free 2d CFT. 

This suggests that ambitwistor string theories can be used to study perturbative QFT on curved backgrounds. There are promising signs that this could be true: in the special case of four-dimensions, twistor string formulae for gravitational scattering amplitudes have a natural generalization to gauged supergravity on AdS$_4$~\cite{Adamo:2015ina}. These formulae pass several consistency checks which indicate that they encode tree-level physical observables in AdS$_4$ (at least up to boundary terms), but so far a direct link to standard expressions in general space-time dimension is missing.  

\medskip

In this paper, we quantize ambitwistor strings on plane wave backgrounds, showing that they encode the correct spectra of perturbations in terms of explicit vertex operators as well as the correct space-time interactions by computing 3-point functions. Our focus is on the type II and heterotic ambitwistor strings, which describe gravitational and gauge theoretic degrees of freedom on space-time, respectively. We study two simple examples of curved backgrounds in a RNS worldsheet formalism for the ambitwistor string: vacuum plane wave metrics (type II model) and plane wave abelian gauge fields (heterotic model). In both cases, the BRST cohomology is shown to correctly encode linearized perturbations around the non-linear background fields, and explicit vertex operators for gravitons or gluons are derived. These vertex operators are then used to compute the 3-point worldsheet correlation functions, which match with the known field theory results in each case \cite{Adamo:2017nia}.  Indeed the computations of \cite{Adamo:2017nia} were first executed in order to provide a standard space-time computation against which to check the formulae of this paper. This confirms the utility of ambitwistor strings in the study of perturbative field theory on curved backgrounds: such calculations are impossible in ordinary string theory, even in the $\alpha'\rightarrow 0$ limit.

After a brief review of the type II and heterotic ambitwistor strings in flat space, we discuss how these models can be defined on curved background fields in Section~\ref{WSM}. Our focus is on background metric fields for the type II model and abelian background gauge fields for the heterotic model. In each case, quantum consistency of the model is equivalent to the usual field equations for the background. Section~\ref{grPW} considers the type II model on a vacuum plane wave metric background, and Section~\ref{gtPW} considers the heterotic model on a plane wave gauge field background. In both cases, the worldsheet theory is anomaly-free because the backgrounds solve the (vacuum) equations of motion. We construct vertex operators corresponding to gravitons and gluons, respectively, and compute their 3-point functions. These are seen to reproduce the known formulae for graviton and gluon scattering on gravitational and gauge field plane wave backgrounds.


\section{Worldsheet Models}
\label{WSM}

Ambitwistor string theories are worldsheet models whose spectra contain only massless degrees of freedom. Our focus will be on those models whose spectra include ordinary, massless supergravity and gauge theory; these are known as the type II and heterotic ambitwistor strings, respectively. After a brief review of these models on flat backgrounds, we describe how they can be coupled to curved background gravitational and gauge fields.


\subsection{Ambitwistor strings in flat space}

For flat backgrounds, ambitwistor strings are given by constrained chiral CFTs in two dimensions, governing holomorphic maps from a Riemann surface $\Sigma$ to ambitwistor space, the space of complex null geodesics in $d$-dimensional Minkowski space~\cite{Lebrun:1983pa}. There is a small zoo of these ambitwistor strings~\cite{Ohmori:2015sha,Casali:2015vta,Azevedo:2017lkz}, but we will be interested in two particular models: the \emph{type II} and \emph{heterotic} ambitwistor strings, which were introduced in~\cite{Mason:2013sva}.

The type II ambitwistor string is described by the worldsheet action (in conformal gauge):
\begin{align}
\label{IIAction}
S=\frac{1}{2\pi}\int_\Sigma P_\mu\, \bar{\partial}X^\mu + \tilde{\psi}_\mu\, \bar{\partial}\psi^\mu - \frac{e}{2}\,\eta ^{\mu\nu} P_\mu P_\nu +\tilde\chi\, \psi^\mu P_\mu + \chi\, \eta^{\mu\nu} \tilde{\psi}_\mu P_\nu\,,
\end{align}
with the worldsheet matter fields $\{P_{\mu}, X^{\mu}, \tilde{\psi}_{\mu}, \psi^{\mu}\}$ having holomorphic conformal weight $\{1,0,\frac{1}{2}, \frac{1}{2}\}$, respectively.\footnote{This form of the type II ambitwistor string, given in~\cite{Adamo:2014wea}, combines the two real worldsheet Majorana fermion systems of the original formulation~\cite{Mason:2013sva} into a single complex fermion system.} The $PX$-system has bosonic statistics, while the $\tilde{\psi}\psi$-system is fermionic. In other words,
\be
P_\mu \in\Omega^0(\Sigma,K_\Sigma)\,, \qquad X^{\mu}\in\Omega^0(\Sigma)\,, \qquad \tilde{\psi}_{\mu},\psi^{\nu}\in\Pi\Omega^0(\Sigma, K^{1/2}_\Sigma)\,.
\ee
The gauge fields $e$, $\tilde{\chi}$, $\chi$ act as Lagrange multipliers, enforcing the constraints $P^2=0$, $\psi\cdot P=0=\tilde{\psi}\cdot P$, and carry non-trivial conformal weights:
\be
e\in\Omega^{0,1}(\Sigma, T_{\Sigma})\,, \qquad \tilde\chi,\chi \in\Pi\Omega^{0,1}(\Sigma,T^{1/2}_\Sigma)\,.
\ee
The constraints imposed by these Lagrange multipliers are conjugate to the gauge transformations
\begin{align}
 \delta X^{\mu}=\alpha\,\eta^{\mu\nu}P_{\nu} -\epsilon\,\eta^{\mu\nu}\tilde{\psi}_{\nu}-\tilde{\epsilon}\,\psi^{\mu}\,, \qquad \delta P_{\mu}=0\,, \nonumber \\
 \delta \psi^{\mu}=\epsilon\,\eta^{\mu\nu}P_{\nu}\,, \qquad \delta\tilde{\psi}_{\mu}=\tilde{\epsilon}\,P_{\mu}\,, \nonumber \\
 \delta e= \dbar\alpha +2(\chi\,\tilde{\epsilon}+\tilde{\chi}\epsilon)\,, \qquad \delta\chi=\dbar\epsilon\,, \qquad \delta\tilde{\chi}=\dbar\tilde{\epsilon}\,, \nonumber
\end{align}
where $\alpha$ is a bosonic gauge parameter of holomorphic conformal weight $-1$ and $\epsilon,\tilde{\epsilon}$ are fermionic gauge parameters of holomorphic conformal weight $-\frac{1}{2}$. These gauge symmetries effectively reduce the target space to (super) ambitwistor space.

The gauge freedoms can be used to set $e=\chi=\tilde{\chi}=0$ via the standard Fadeev-Popov procedure. The resulting gauge-fixed action is manifestly free:
\be\label{gfIIAct}
S=\frac{1}{2\pi}\int_{\Sigma} P_\mu\, \bar{\partial}X^\mu + \tilde{\psi}_\mu\, \bar{\partial}\psi^\mu + b\,\dbar c+\tilde{b}\,\dbar\tilde{c}+\beta\,\dbar\gamma +\tilde{\beta}\,\dbar\tilde{\gamma}\,.
\ee
The $c$-ghost, a fermionic field of conformal weight $(-1,0)$, is associated with holomorphic reparametrization invariance on the worldsheet, and $\tilde{c}$ (with the same quantum numbers as $c$) is associated with the gauge transformations generated by the $P^2=0$ constraint. The bosonic ghosts $\gamma,\tilde{\gamma}$ are both left-moving, with conformal weight $(-\frac{1}{2},0)$, and are associated with the gauge transformations generated by the constraints $\psi\cdot P=0=\tilde{\psi}\cdot P$. 

The BRST-charge resulting from this gauge fixing procedure is
\be\label{IIBRST}
Q=\oint c\,T+bc\,\partial c +\frac{\tilde{c}}{2}\,\eta^{\mu\nu}P_{\mu}P_{\nu} + \tilde{\gamma}\,\psi^{\mu} P_{\mu}+\gamma\,\eta^{\mu\nu}\tilde{\psi}_{\mu} P_{\nu} -2\gamma\tilde{\gamma}\tilde{b}\,,
\ee
where $T$ is the holomorphic stress tensor. Using the free OPEs associated with \eqref{gfIIAct}
\be\label{flatOPEs}
X^{\mu}(z)\,P_{\nu}(w)\sim \frac{\delta^{\mu}_{\nu}}{z-w}\,, \qquad \psi^{\mu}(z)\,\tilde{\psi}_{\nu}(w)\sim \frac{\delta^{\mu}_{\nu}}{z-w}\,,
\ee
and likewise for the ghost fields, it is straightforward to calculate any possible anomalies. Indeed, one finds
\be\label{IIQ2}
Q^2=(d-10)\,\frac{c\,\partial^{3} c}{4}\,,
\ee
so the only anomaly is the central charge, which is eliminated in the critical target dimension $d=10$. As long as the worldsheet is genus zero, $\Sigma\cong\CP^1$, this conformal anomaly will not affect the computation of worldsheet correlation functions. So from the point of view of scattering amplitudes, the type II ambitwistor string is well-defined on Minkowski space of any dimension at genus zero.

Using the BRST operator \eqref{IIBRST}, one can investigate the spectrum of the model, which is in one-to-one correspondence with that of type II supergravity~\cite{Mason:2013sva,Adamo:2013tsa}. For instance, it is easy to see that in the NS-NS sector, fixed vertex operators of the form
\be\label{IIfixed}
c\,\tilde{c}\,\delta(\gamma)\,\delta(\tilde{\gamma})\,\tilde{\epsilon}^{\mu}\epsilon_{\nu}\,\tilde{\psi}_{\mu}\,\psi^{\nu}\,\e^{\im\, k\cdot X}\,,
\ee
are $Q$-closed provided $k^2=k\cdot\epsilon=k\cdot\tilde{\epsilon}=0$. The symmetric, antisymmetric and trace parts of $\tilde{\epsilon}^{\mu}\epsilon_{\nu}$ encode the massless gravitons, $B$-fields and dilaton of type II supergravity. A key feature of the ambitwistor string is that the $n$-point sphere correlation functions of these vertex operators, along with their picture number zero descendants, are equal to the CHY formulae for the tree-level scattering amplitudes of supergravity. 

So to summarize: the type II ambitwistor string on a Minkowski background has the spectrum of massless type II supergravity, is well-defined up to a conformal anomaly (which is irrelevant at genus zero), and produces the tree-level S-matrix of supergravity perturbatively around Minkowski space in terms of worldsheet correlation functions.

\medskip

The heterotic ambitwistor string, as its name suggests, is obtained by replacing the complex fermion system of the type II model with a single real fermion system while simultaneously adding a holomorphic worldsheet current algebra. In Minkowski space, the worldsheet action in conformal gauge is given by
\be\label{HetAct}
S=\frac{1}{2\pi}\int_{\Sigma}P_\mu\, \bar{\partial}X^\mu + \Psi_\mu\, \bar{\partial}\Psi^\mu -\frac{e}{2} \eta^{\mu\nu}\,P_{\mu}P_{\nu}+\chi\,\Psi\cdot P +\mathcal{L}_{C}\,,
\ee
where $\Psi^{\mu}$ are fermionic with holomorphic conformal weight $\frac{1}{2}$, and $\cL_{C}$ is the Lagrangian for a holomorphic worldsheet current algebra. As before, holomorphic reparametrization invariance and the gauge freedoms associated with the constraints $P^2=0$ and $\Psi\cdot P=0$ can be used to set $e=\chi=0$. This results in a gauge fixed action
\be\label{gfhAct}
S=\frac{1}{2\pi}\int_{\Sigma}P_\mu\, \bar{\partial}X^\mu + \Psi_\mu\, \bar{\partial}\Psi^\mu + b\,\dbar c+\tilde{b}\,\dbar\tilde{c}+\beta\,\dbar\gamma + \cL_{C},
\ee
and BRST charge
\be\label{HetBRST}
Q=\oint c\,T+bc\,\partial c +\frac{\tilde{c}}{2}\,\eta^{\mu\nu}P_{\mu}P_{\nu} + \gamma\,\Psi^{\mu} P_{\mu}-\frac{\tilde{b}}{2}\,\gamma\,\gamma\,,
\ee
where the ghost systems have the same statistics and quantum numbers as before. 

The only obstruction to $Q^2=0$ for the heterotic model is again given by the central charge, which is $\frac{5}{2}d-41 +\mathfrak{c}$, with $\mathfrak{c}$ the central charge of the worldsheet current algebra. So for any fixed $d\leq 16$, this anomaly can be eliminated by choosing the worldsheet current algebra appropriately. However, at genus zero the conformal anomaly is practically irrelevant.

In the gauge theory sector, the spectrum of the heterotic model agrees with that of $\cN=1$ super-Yang-Mills theory. Take the fixed NS sector vertex operators
\be\label{hetfixed}
c\,\tilde{c}\,\delta(\gamma)\,\epsilon\cdot\Psi\,j^{\sa}\,\sT^{\sa}\,\e^{\im\,k\cdot X}\,,
\ee
with $j^{\sa}$ the worldsheet current for some Lie algebra $\mathfrak{g}$ and $\sT^{\sa}$ the generators, where the sans-serif Roman indices $\sa,\ldots=1,\ldots, \dim \mathfrak{g}$. These vertex operators are $Q$-closed provided $k^2=k\cdot\epsilon=0$, and therefore represent gluons. Correlation functions of such vertex operators (and their descendants) at genus zero lead to the CHY expressions for the tree-level scattering amplitudes of Yang-Mills theory in $d$-dimensional Minkowski space. 

The gravitational sector of the heterotic ambitwistor string corresponds to a certain non-unitary $R^2$ supergravity~\cite{Azevedo:2017lkz}. Since our considerations will be entirely at genus zero, we can consistently project out these modes, focusing only on the gauge theory sector corresponding to Yang-Mills theory from now on.

%


\subsection{Type II model on a curved background metric}

In~\cite{Adamo:2014wea} it was shown how to couple the type II ambitwistor string to curved background fields from the NS-NS supergravity sector (metric, $B$-field and dilaton). Here, we review this construction with only the background metric turned on, as the $B$-field and dilaton will not be relevant for our later calculations.

Let $g_{\mu\nu}$ be the space-time metric. The curved space analogue of the gauge-fixed worldsheet action \eqref{gfIIAct} is
\be\label{IIcurv1}
S=\frac{1}{2\pi}\int_{\Sigma} P_\mu\, \bar{\partial}X^\mu + \tilde{\psi}_\mu\, \bar{\partial}\psi^\mu +\tilde{\psi}_{\mu}\,\psi^{\nu}\,\Gamma^{\mu}_{\nu\rho}\,\dbar X^{\rho} + b\,\dbar c+\tilde{b}\,\dbar\tilde{c}+\beta\,\dbar\gamma +\tilde{\beta}\,\dbar\tilde{\gamma}\,,
\ee
where $\Gamma^{\mu}_{\nu\rho}$ are the Christoffel symbols for the Levi-Civita connection of $g_{\mu\nu}$. At first, this may not seem very promising: the connection term (required to ensure space-time covariance of the worldsheet action) couples the fermions to $X^{\mu}$ non-polynomially. However, it was observed in~\cite{Adamo:2014wea} that the field redefinition
\be\label{Pidef}
P_{\mu}\rightarrow \Pi_{\mu}:=P_{\mu}+\tilde{\psi}_{\rho}\,\psi^{\nu}\,\Gamma^{\rho}_{\mu\nu}\,,
\ee
leaves a free worldsheet action; the price for this simplification is that the new field $\Pi_{\mu}$ does not transform covariantly under space-time diffeomorphisms. This is a small price to pay for a manifestly solvable 2d CFT on \emph{any} curved target space-time, though.

After taking into account certain subtleties associated with worldsheet reparametrization invariance (see~\cite{Adamo:2014wea} for details), the worldsheet action for the type II model on a curved target space metric is:
\be\label{IIcurv2}
S=\frac{1}{2\pi}\int_{\Sigma} \Pi_\mu\, \bar{\partial}X^\mu + \tilde{\psi}_\mu\, \bar{\partial}\psi^\mu + b\,\dbar c+\tilde{b}\,\dbar\tilde{c}+\beta\,\dbar\gamma +\tilde{\beta}\,\dbar\tilde{\gamma} +\frac{R_{\Sigma}}{4}\,\log\left(\sqrt{g}\right)\,,
\ee
where $R_{\Sigma}$ is the scalar curvature of the worldsheet and $g$ is the (absolute value of the) determinant of the metric. Locally, $R_{\Sigma}$ can always be taken to vanish, so the final term in the action does not affect the worldsheet OPEs of the model, which are the same as in flat space with $\Pi_\mu$ playing the role of $P_\mu$.

Associated with this gauge-fixed action is a curved version of the BRST charge \eqref{IIBRST}, taking the form:
\be\label{IIcQ}
Q=\oint c\,T+bc\,\partial c +\frac{\tilde{c}}{2}\,\cH + \tilde{\gamma}\,\cG +\gamma\,\tilde{\cG}-2\gamma\tilde{\gamma}\tilde{b}\,,
\ee
where the currents $\cG$, $\tilde{\cG}$ and $\cH$ generalize $\psi\cdot P$, $\tilde{\psi}\cdot P$ and $P^2$ to curved space, respectively. They are given by 
\be\label{GeneralG}
\mathcal{G}=\psi^\mu\,\Pi_\mu + \partial(\psi^\mu\Gamma^\kappa{}_{\mu\kappa})\,,
\ee
\be\label{GeneralGbar}
 \tilde{\mathcal{G}}=g^{\mu\nu}\,\tilde\psi_\nu\,\left(\Pi_\mu-\Gamma^\kappa{}_{\mu\lambda}\,\tilde\psi_\kappa\psi^\lambda\right) - g^{\mu\nu}\,\partial(\tilde\psi_\kappa\Gamma^\kappa{}_{\mu\nu})\,,
\ee
and
\begin{align}
\begin{split}
\label{GeneralH}
\mathcal{H}=& g^{\mu\nu}\left(\Pi_\mu-\Gamma^\kappa{}_{\mu\lambda}\tilde\psi_\kappa\psi^\lambda\right)\left(\Pi_\nu-\Gamma^\kappa{}_{\nu\lambda}\tilde\psi_\kappa\psi^\lambda\right) -\frac{1}{2}R^{\kappa\lambda}{}_{\mu\nu}\,\tilde\psi_\kappa\tilde\psi_\lambda\psi^\mu\psi^\nu  
\\ 
& 
-g^{\mu\nu}\partial \left( \Gamma^\rho_{\mu\nu} \Pi_\rho \right) 
-\frac{1}{2}\partial^2(g^{\mu\nu})\partial_\mu\partial_\nu \log(\sqrt{g}) -\tilde\psi_\kappa\partial\psi^{\lambda}\, g^{\mu\nu}\partial_\lambda\Gamma^\kappa{}_{\mu\nu}
\\
&-\frac{1}{4}\partial(g^{\mu\nu})\,\partial(\partial_\mu\partial_\nu\log(\sqrt{g})) -\partial\left[\frac{\partial g^{\mu\nu}}{2}\left(\Gamma^\sigma_{\mu\nu}\Gamma^\rho_{\sigma\rho} - \partial_\sigma \Gamma^\sigma_{\mu \nu} \right) + g^{\mu\nu} \Gamma^\rho_{\mu \sigma} \partial(\Gamma^\sigma_{\nu\rho})\right]\,.
\end{split}
\end{align}
Though they don't look covariant, it was shown in~\cite{Adamo:2014wea} that these currents do indeed transform correctly under diffeomorphisms in the full quantum worldsheet theory.

Using the BRST charge and free OPEs of the worldsheet action, the anomalies of the type II model on a curved background can be computed exactly. As in flat space, there is a conformal anomaly; remarkably, this anomaly is unaffected by the curved space-time metric. In particular, it vanishes in $d=10$ space-time dimensions and can be ignored at genus zero for the purposes of calculating scattering amplitudes. However, there are also new anomalies which emerge only when the space-time metric is non-trivial. These anomalies are related to the current algebra between $\cG$, $\tilde{\cG}$ and $\cH$; they vanish if and only if the OPEs between the currents obey
\be\label{IIcurv3}
\cG(z)\,\cG(w)\sim 0 \sim \tilde{\cG}(z)\,\tilde{\cG}(w)\,, \qquad \cG(z)\,\tilde{\cG}(w)\sim \frac{\cH}{z-w}\,.
\ee
Remarkably, this happens precisely when the space-time metric obeys the vacuum Einstein equations: $R_{\mu\nu}=0$.

Hence, the type II model coupled to a curved space-time metric is anomaly free (ignoring the conformal anomaly) if and only if this metric solves the vacuum Einstein equations. This is analogous to the statement that ordinary string theory is anomaly free at lowest order in $\alpha'$ if and only if the Einstein equations hold~\cite{Callan:1985ia,Fradkin:1985ys,Banks:1986fu}. But unlike ordinary string theory on a curved background, where the worldsheet action is a complicated interacting 2d CFT necessitating perturbation theory to determine anomalies, the ambitwistor string remains solvable, the anomaly is obtained exactly, and no perturbative expansion is required.




\subsection{Heterotic model on a gauge field background}
\label{Hcb}

Let $\sA_{\mu}$ be a background gauge field on space-time, taking values in the adjoint of some Lie algebra $\mathfrak{g}$. For simplicity, we assume that $\sA$ is in the Cartan of the gauge group, so that it may be treated as an abelian gauge field.\footnote{The case of a general background gauge field is more complicated; we hope to address this general case in future work.} The idea is now to couple the heterotic ambitwistor string to this abelian background gauge field.

This coupling occurs through the worldsheet current algebra, which we assume to have level zero. For example, consider a realization of the worldsheet current algebra for SO$(N)$ in terms of $N$ real fermions
\be\label{wca1}
S_{C}=\frac{1}{2\pi}\int_{\Sigma} \rho^{A}\,\dbar\rho^{A}\,,
\ee
where $A$ runs over the fundamental representation of $\mathfrak{so}(N)$. The sum over repeated indices is achieved implicitly with the Killing form, $\delta^{AB}$. The conformal weight $(1,0)$ worldsheet current, which takes values in the adjoint of the gauge group is $j^{\sa}=(\mathsf{T}^{\sa})_{AB}\rho^{A}\rho^{B}$.  Coupling to the background gauge field is then given by:
\be\label{wca2}
S_{C}\rightarrow \frac{1}{2\pi}\int_{\Sigma} \rho^{A}\,\dbar\rho^{A} + \sA^{\sa}_{\mu}\,j^{\sa}\,\dbar X^{\mu}\,,
\ee
with summation over all Lie algebra and space-time indices assumed. 

It is clear that the coupling to the background field can be absorbed into a field redefinition of $P_{\mu}$, as in the type II model. In the case of \eqref{wca2}, this field redefinition is
\be\label{hfrdef}
P_{\mu}\rightarrow \Pi_{\mu}:= P_{\mu}+ \sA_{\mu}^{\sa}\,j^{\sa}\,.
\ee
As in the type II model, the new field $\Pi_{\mu}$ does not behave covariantly under gauge transformations, but the resulting gauge-fixed worldsheet action is free:
\be\label{hetcurv1}
S=\frac{1}{2\pi}\int_{\Sigma} \Pi_\mu\, \bar{\partial}X^\mu + \Psi_\mu\, \bar{\partial}\Psi^\mu + b\,\dbar c+\tilde{b}\,\dbar\tilde{c}+\beta\,\dbar\gamma + \cL_{C}\,,
\ee
with $\cL_{C}$ the same worldsheet current algebra as in the flat background. To this worldsheet action we associate a BRST charge
\be\label{hetcQ}
Q=\oint c\,T+bc\,\partial c +\frac{\tilde{c}}{2}\,\sH + \gamma\,\sG-\frac{\tilde{b}}{2}\,\gamma\,\gamma\,,
\ee
where $\sG$ and $\sH$ are the analogues of the currents $\Psi\cdot P$ and $P^2$ in the presence of the background gauge field. They are given by
\be\label{hetG}
\sG=\Psi^{\mu}\left(\Pi_{\mu}-\sA^{\sa}_{\mu}\,j^{\sa}\right)\,,
\ee
and
\be\label{hetH}
\sH=\eta^{\mu\nu}\left(\Pi_{\mu}-\sA^{\sa}_{\mu}\,j^{\sa}\right) \left(\Pi_{\nu}-\sA^{\sa}_{\nu}\,j^{\sa}\right)+\Psi^{\mu}\,\Psi^{\nu}\, \mathsf{F}^{\sa}_{\mu\nu}\,j^{\sa} -\partial\left(\partial^{\mu}\sA^{\sa}_{\mu} j^{\sa}\right)\,,
\ee
with $\mathsf{F}_{\mu\nu}=\partial_{\mu}\sA_{\nu}-\partial_{\nu}\sA_{\mu}$ the field strength of the abelian background gauge field.

Using the free worldsheet OPEs of \eqref{hetcurv1} and the level zero current algebra identity
\be\label{lzeroOPE}
j^{\sa}(z)\,j^{\mathsf{b}}(w)\sim \frac{f^{\sa \mathsf{b}\mathsf{c}}\,j^{\mathsf{c}}}{z-w}\,,
\ee
where $f^{\sa \mathsf{b}\mathsf{c}}$ are the structure constants of the gauge group, the anomalies of the model can be computed explicitly. As in the type II model, there are two kinds of anomaly: a conformal anomaly which is the same as in the flat background, and a current algebra anomaly associated with $\sG$ and $\sH$. The former will not affect genus zero scattering amplitudes (as in flat space), but the latter must be eliminated. It is easy to see that $Q^2=0$ up to the conformal anomaly if and only if the curved background currents obey:
\be\label{hetcuralg}
\sG(z)\,\sG(w)\sim \frac{\sH}{z-w}\,, \qquad \sG(z)\,\sH(w)\sim 0\,.
\ee
The first identity is trivially obeyed, and a direct calculation shows that
\be\label{hetcurv2}
\sG(z)\,\sH(w)\sim -3\, \frac{\Psi^{\nu}\,\partial^{\mu}\mathsf{F}^{\sa}_{\mu\nu}\,j^{\sa}}{(z-w)^2}-\frac{\partial(\Psi^{\nu}\partial^{\mu}\mathsf{F}^{\sa}_{\mu\nu}\, j^{\sa})}{z-w} -\frac{\Psi^{\mu}\Psi^{\nu}\Psi^{\sigma}\, \partial_{\mu}\mathsf{F}^{\sa}_{\nu\sigma}\, j^{\sa}}{z-w}\,.
\ee
So the anomaly vanishes if and only if the background gauge field obeys the Bianchi identity and Maxwell equations:
\be\label{hetcurv3}
\partial_{[\mu}\mathsf{F}^{\sa}_{\nu\sigma]}=0\,, \qquad \partial^{\mu}\mathsf{F}^{\sa}_{\mu\nu}=0\,.
\ee
Once more, we note that these field equations emerge from an \emph{exact} anomaly calculation in a \emph{free} 2d CFT, unlike the analogous calculations in the heterotic~\cite{Callan:1985ia} or type I~\cite{Abouelsaood:1986gd} superstring.


\section{Type II Model on a Gravitational Plane Wave}
\label{grPW}

We now turn to a class of backgrounds which are among the simplest non-trivial solutions to the vacuum Einstein equations. When discussing gravitational waves, one often thinks about linear perturbations of a fixed background space-time as first described by Einstein~\cite{einstein1916}. However, there also are exact, non-linear plane wave solutions to the Einstein equations. These have been known for over ninety years and studied in great detail, see e.g.~\cite{Baldwin:1926,Ehlers:1962zz,griffiths1991colliding,Stephani:2003tm,Blau:2011}. Due to the vanishing of their higher-curvature invariants, it has long been known that certain plane wave metrics solve the vanishing beta-functional conditions for string theory to all orders in $\alpha'$~\cite{Amati:1988sa,Horowitz:1989bv}. Yet it has proven difficult to derive vertex operators for string theory on a plane wave background or indeed compute scattering amplitudes in such space-times.\footnote{A notable exception is~\cite{Jofre:1993hd}, where candidate tachyon and graviton vertex operators are constructed in the bosonic string on a plane wave.}

In this section, we study the type II ambitwistor string on vacuum plane wave space-times. Unlike ordinary string theory, the worldsheet OPEs remain free and computing vertex operators and worldsheet correlation functions is tractable. After a brief review of gravitational plane waves and their scattering theory, we construct the vertex operators (in fixed and descended pictures) corresponding to gravitons on this background using the BRST cohomology of the worldsheet theory.  We go on to show that the 3-point correlation functions on the Riemann sphere reproduce the known result for 3-point graviton amplitudes on a plane wave space-time.


\subsection{Plane wave metrics \& Scattering}

Two coordinate systems are most commonly used to describe gravitational plane wave metrics: Einstein-Rosen coordinates~\cite{Einstein:1937qu} or Brinkmann coordinates~\cite{Brinkmann:1925fr}. The line element in Einstein-Rosen coordinates is $$
\d s^2 = 2\,\d U\, \d V - \gamma_{ij}(U) \d y^i \d y^j,
$$
 with the Roman indices $1\le i,j \le d-2$. We define $ \gamma^{ij}$ to be the inverse of the $(d-2)\times(d-2)$ block $\gamma_{ij}$. Although these coordinates manifest $d-1$ of the $2d-3$ Killing vectors possessed by plane wave metrics, they develop singularities~\cite{Penrose:1965rx,Bondi:1989vm} and the vacuum equations are given by a second-order ODE on $\gamma_{ij}$:
\be\label{ERRicci}
\left(\ddot{\gamma}_{ij} + \frac{1}{2}\dot{\gamma}_{ik} \gamma^{kl} \dot{\gamma}_{lj} \right) \gamma^{ij}=0\,,
\ee
where a dot denotes $\partial_U$. 

In the Brinkmann coordinate system, there is only a single non-trivial component in the plane wave metric:
\be\label{Brink1}
\d s^2 = 2\,\d u\,\d v -H(u,\mathbf{x})\,\d u^2 - \delta_{ab}\,\d x^{a}\,\d x^{b}\,,
\ee
where $1\leq a,b\leq d-2$ are flat indices and the non-trivial metric function $H$ is quadratic in $x^a$:
\be\label{Brink2}
H(u,\mathbf{x})=H_{ab}(u)\,x^{a}\,x^b\,.
\ee
This coordinate system has the advantages of being global and encoding the curvature directly through $H$. Indeed, the only non-trivial Riemann tensor components are
\be\label{Rcurv}
R^{a}_{\:\:ubu}=-H^{a}_{b}(u)\,,
\ee
and the only non-zero Christoffel symbols are
\be\label{Christoffel}
\Gamma^{a}_{uu}=-H_{ab}(u)\,x^{b}\,, \quad \Gamma^{v}_{ua}=-H_{ab}(u)\,x^{b}\,, \quad \Gamma^{v}_{uu}=-\frac{\dot{H}(u,\mathbf{x})}{2}\,.
\ee
The transformation from Einstein-Rosen to Brinkmann coordinates is given by
\begin{align}\label{diffeo}
\begin{split}
U& = u\,,
\\
V &= v+ \frac{1}{2} \dot{E}^i_a\, E_{b\,i}\, x^a x^b\,,
\\
y^i &= E^i_a \, x^a\,,
\end{split}
\end{align}
where the vielbein $E^a_i$ is defined by $\gamma_{ij} = E^a_i E^b_j \delta_{ab}$ and satisfies
\be\label{vielbein}
\ddot{E}_{a\,i}=H_{ab}\,E^{b}_{i}\,, \qquad \dot{E}^{a}_{[i}\,E_{|a|\,j]}=0\,. 
\ee
The combination
\be\label{shear}
\sigma_{ab}=\dot{E}^i_a\,E_{b\,i}\,,
\ee
encodes the expansion and shear of the $\partial_U$ null geodesic congruence in its trace and trace-free parts, respectively, and often appears in the study of perturbative gravity on a background plane wave.

In the following, we mainly work in Brinkmann coordinates; however, the availability of Einstein-Rosen coordinates will be key when solving the linearised Einstein equations on this background.

\medskip

One may now ask: is there a sensible notion of S-matrix for massless particles propagating on a plane wave space-time? On a general curved background, it is not at all clear how to set up scattering problems in a meaningful way. Recall that the usual S-matrix is defined via in- and out-states on a flat (or at least asymptotically flat) background. Furthermore, it is not guaranteed that time evolution will be unitary. In addition, there might be particle creation in the space-time, as in the case of Hawking radiation from black holes.
\begin{figure}[t]
\centering
\includegraphics[scale=.6]{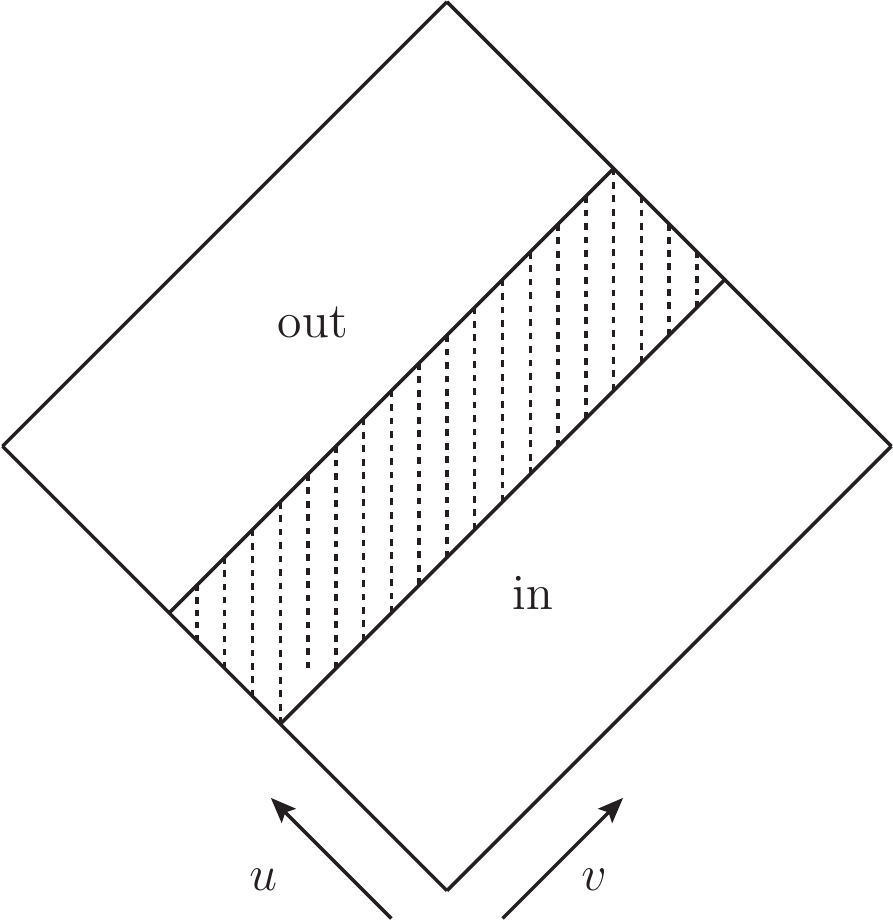}
\caption{A two dimensional Penrose diagram of a sandwich plane wave, ignoring the spatial directions. While the far past and future are flat, the grey region defined by $u_1\le u \le u_2$ has non-vanishing curvature.}
\label{Sandwich}
\end{figure}

The first issue is solved easily in the case of a plane wave space-time by considering a so-called \emph{sandwich plane wave}~\cite{Bondi:1958aj}, where $H_{ab}(u)$ has compact support in $u_1\le u \le u_2$, see Figure~\ref{Sandwich}. These sandwich plane waves have flat in- and out-regions for $u<u_1$ and $u>u_2$, respectively, so in- and out-states can still be defined. The problem of unitary time evolution is more involved. Penrose proved that plane wave space-times never admit a Cauchy surface~\cite{Penrose:1965rx} and hence are not globally hyperbolic. Usually the natural inner product between states of free theories is defined by integration over a Cauchy surface. Here the situation is rescued by defining the inner product by integration over a surface of constant $u$. This can then be used to show that time evolution is actually unitary and also that there is no particle creation. These properties for scalar free fields on sandwich plane waves have been known for some time~\cite{Gibbons:1975jb,Garriga:1990dp}, and were recently generalised to the cases of spin one and two~\cite{Adamo:2017nia}.

Another potential issue when computing scattering amplitudes is that momentum eigenstates from the in-region will not evolve to momentum eigenstates in the out-region. This is essentially a consequence of the memory effect: an in state, or momentum eigenstate in the in-region, has a planar wavefront in the in-region, but in the out region its wavefront could be a plane, sphere, cylinder, cone, torus or certain types of Dupin cyclides~\cite{Friedlander:1975eqa,Ward:1987ws}. This means that scattering mixtures of in- and out-states substantially complicates the integrals arising in amplitude calculations, even at 3-points~\cite{Adamo:2017nia}. This subtlety can be avoided by considering the scattering of only a single type of state (i.e., all in- or all out-states), for which many integrals simplify. Since the `tree-level integrands' of the amplitudes are independent of the state configuration and carry all of the information required to obtain the actual amplitudes, we will consider only the scattering of one type of state in the following.


\subsection{Worldsheet model and Vertex operators}

As solutions of the vacuum Einstein equations ($H_{a}^{a}(u)=0$ in Brinkmann coordinates), vacuum plane wave metrics are admissible backgrounds for the type II ambitwistor string in the sense that $Q^2=0$, up to a conformal anomaly which can be killed by setting $d=10$, or ignored for our purposes at genus zero. However, a remarkable simplification occurs in the functional form of the BRST charge for any plane wave background in Brinkmann coordinates: all quantum corrections to the currents $\cG$, $\tilde{\cG}$ and $\cH$ vanish! Indeed, a direct calculation from \eqref{GeneralG} -- \eqref{GeneralH} using the Christoffel symbols \eqref{Christoffel} leads to:
\begin{align}
\begin{split}
\label{BrinkmannG}
\mathcal{G} &= \psi^\mu\, \Pi_\mu\,,
\\
\tilde{\mathcal{G}} &= g^{\mu \sigma}\, \tilde{\psi}_\mu\, \Pi_\sigma + 2H^{a}_{b}\,x^{b}\, \tilde{\psi}_v \,\tilde{\psi}_a\,  \psi^u\,,
\end{split}
\end{align}
\begin{align}
\begin{split}
\label{BrinkmannH}
\mathcal{H} &= g^{\mu\sigma}\,\Pi_{\mu}\,\Pi_{\sigma} + \Pi_v \left(2 H^{a}_{b}\,x^{b}\, \tilde{\psi}_a  \psi^u +2 H_{ab}\,x^{b}\,\tilde{\psi}_{v}\psi^{a}+ \dot{H}\, \tilde{\psi}_v \psi^u \right)
\\
& \qquad  -2\Pi_{a}\,H^{a}_{b}\,x^{b}\,\tilde{\psi}_{v}\psi^{u} +2H^{a}_{b}\,\tilde{\psi}_{a} \tilde{\psi}_v\, \psi^b \psi^u\,.
\end{split}
\end{align}
In particular, all terms proportional to worldsheet derivatives in $\cG$, $\tilde{\cG}$ and $\cH$ vanish in Brinkmann coordinates for the plane wave background.

With a concrete expression for the BRST charge and free worldsheet OPEs, we are now in a position to determine vertex operators lying in the BRST cohomology of the model. For simplicity, we focus on graviton-type vertex operators in the NS-NS sector; the generalization to other states is straightforward. Consider an ansatz for a vertex operator in the fixed picture:
\begin{align}
\begin{split}
\label{FixedVOBrinkmann}
V= c\, \tilde{c}\, \delta(\gamma)\, \delta(\tilde{\gamma})\, \tilde{\psi}_\mu\, \psi^\sigma h^\mu_\sigma\,,
\end{split}
\end{align}
where $h_{\nu\sigma}=h_{\sigma\nu}$ is a function of the space-time coordinates $X^{\mu}$ only, and has vanishing worldsheet conformal weight: $h_{\mu\sigma}$ will be a metric perturbation (i.e., a graviton) on the plane wave background. We will find that it is consistent to further impose the gauge conditions  that the $v$-components and trace of $h$ should vanish $h_{v \mu}=0=h^{\mu}_{\mu}$.

Since $V$ has balanced conformal weight, the stress tensor part of the BRST charge acting on this vertex operator vanishes. The only non-trivial conditions for BRST-closure of $V$ arise from the currents $\cG$, $\tilde{\cG}$ and $\cH$. It is easy to see that $QV=0$ if and only if the OPEs
\be\label{IIQclose}
\cG(z)\,\tilde{\psi}_{\mu}\psi^{\sigma} h^{\mu}_{\sigma}(w)\,, \qquad \tilde{\cG}(z)\,\tilde{\psi}_{\mu}\psi^{\sigma} h^{\mu}_{\sigma}(w)\,, \qquad \cH(z)\,\tilde{\psi}_{\mu}\psi^{\sigma} h^{\mu}_{\sigma}(w)\,,
\ee
have only simple pole singularities. A straightforward calculation using the free worldsheet OPEs reveals that:
\begin{align}
\begin{split}
\label{DoublePolesFermion}
\cG(z)\,\tilde{\psi}_{\mu}\psi^{\sigma} h^{\mu}_{\sigma}(w) &\sim  -\frac{\partial_{\mu}h^{\mu}_{\sigma}\psi^{\sigma}}{(z-w)^2} + \frac{1}{z-w}\, (\cdots)\,,
\\
\tilde{\cG}(z)\,\tilde{\psi}_{\mu}\psi^{\sigma} h^{\mu}_{\sigma}(w) &\sim  \frac{g^{\rho\sigma}}{(z-w)^2} \partial_\rho h^\mu_\sigma \tilde{\psi}_\mu + \frac{1}{z-w}\, (\cdots)\,.
\end{split}
\end{align}
Thus, the condition to remove the double poles in these OPEs is $\partial_{\mu}h^{\mu}_{\sigma}=0$. Although this may, at first, appear to be a non-covariant condition, it is in fact equivalent to the de Donder gauge condition $\nabla_\mu h^\mu_\sigma =0$ in Brinkmann coordinates (taking into account the gauge conditions already imposed on $h_{\mu\sigma}$). Hence, as expected from the example of flat space, the fermionic constraints $\cG$, $\tilde{\cG}$ fix a gauge for the metric perturbation. 

We expect that the OPE with $\cH$ will impose equations of motion on $h_{\mu\sigma}$. Computing this OPE, one finds:
\begin{multline}
\cH(z)\,\tilde{\psi}_{\mu}\psi^{\sigma} h^{\mu}_{\sigma}(w)\sim \frac{1}{(z-w)^2} \bigg[g^{\rho\lambda}\partial_\rho \partial_\lambda h^\mu_\sigma\, \tilde{\psi}_\mu \psi^\sigma +2H^{a}_{b}x^{b}\,\partial_{v}h^{\mu}_{a}\,\tilde{\psi}_{\mu}\psi^{u} \\
+2H^{a}_{b}x^{b}\,\partial_{v}h_{\sigma a}\,\tilde{\psi}_{v}\psi^{\sigma} - 2 H^{ab}\,h_{ab}\,\tilde{\psi}_{v}\psi^{u}\bigg] + \frac{1}{z-w}\,(\cdots)\,.
\end{multline}
This imposes the condition
\be\label{linefe}
g^{\rho\lambda}\partial_{\rho}\partial_{\lambda}h_{\mu\sigma}+4\,\delta^{u}_{(\mu} \partial_{|v|} h_{\sigma)a}\,H^{a}_{b}\,x^{b}-2\,\delta^{u}_{\rho}\delta^{u}_{\sigma}\, H^{ab}\, h_{ab}=0\,,
\ee
on $h_{\mu\sigma}$, which is in fact the linearised Einstein equation on the plane wave background.

To summarize, the condition $QV=0$ for a fixed vertex operator $V$ of the form \eqref{FixedVOBrinkmann} imposes that $h_{\mu\sigma}(X)$ is a solution to the linearised Einstein equations in the de Donder gauge which is also trace-free with vanishing $v$-components $h_{v\mu}=0$. These are precisely the desired conditions for a vertex operator representing a target space graviton.

A concrete realization of $h_{\mu\sigma}$, analogous to the momentum eigenstate used in flat space \eqref{IIfixed} can be constructed by a spin-raising procedure~\cite{Mason:1989,Adamo:2017nia} applied to solutions of the scalar wave equation on a plane wave background, first constructed in~\cite{Ward:1987ws}. Key to this construction are solutions to the Hamilton-Jacobi equations of the form:
\be\label{HJsol}
\phi_k=k_0\,v + \frac{k_0}{2}\,\sigma_{ab}\,x^{a}x^{b}+k_i\,E^{i}_{a}\,x^a-\frac{k_{i}\,k_j}{2\,k_0}\,F^{ij}\,,
\ee
where $(k_{0},k_i)$ are $d-1$ constants (which parametrize the nontrivial components of a null momentum), $E^i_a$ is the vielbein appearing in the relationship between Brinkmann and Einstein-Rosen coordinates \eqref{diffeo}, $\sigma_{ab}=\dot{E}^{i}_{a} E_{b\,i}$ and 
\be\label{Fij}
F^{ij}(u):=\int^{u}\d s\,\gamma^{ij}(s) = \int^{u} \d s\,E^{a\,(i}(s)\,E_{a}^{j)}(s)\,.
\ee
The choice of vielbein is not unique: given any $E^a_i$ for a particular plane wave metric, any other vielbein of the form
\be\label{memory}
E^{a\,\mathrm{new}}_{i}=E^{a}_{j}\left(F^{jk}\,b_{ki}+c^{j}_{i}\right)\,,
\ee
for $\mathbf{b}$, $\mathbf{c}$ constant matrices, also represents the same metric. For a sandwich plane wave, two particular choices of boundary condition are relevant:
\be\label{bcs}
\lim_{u\rightarrow\pm\infty}E^{a\,\pm}_{i}=\delta^{a}_{i}\,.
\ee
These correspond to whether $\phi_k$ looks like $k\cdot X$ in the in- or out-regions of the sandwich plane wave; since we will always considering amplitudes in which all external states have the same boundary conditions, we assume that $E^{a}_{i}=E^{a\,-}_{i}$ without loss of generality from now on.

Equipped with the function $\phi_k$, the graviton $h_{\mu\sigma}$ is given by~\cite{Adamo:2017nia}:
\be\label{graviton}
h_{\mu\sigma}\,\d X^\mu\, \d X^\sigma= \left((\varepsilon\cdot \d X)^2-\frac{\im}{k_0}\,\epsilon^{a}\epsilon^{b}\,\sigma_{ab}\, \d u^2\right)\Omega\, \e^{\im \phi_k}
\ee
Here, $\varepsilon_{\mu}$ is a (non-constant) $d$-dimensional polarization vector
\begin{equation}
\label{grpol}
\varepsilon\cdot \d X=\epsilon_a\d x^a + \epsilon^a\left(\frac{k_j}{k_0}E^j_a +  \,\sigma_{ab}\, x^b\right)\d u
\end{equation}
with $\epsilon_a$ a constant $(d-2)$-dimensional null vector. The function $\Omega(u)$ is defined by
\be\label{Omega}
\Omega(u):=|\gamma^{-1}(u)|^{1/4} = |E(u)|^{-\frac{1}{2}}\,.
\ee
It is straightforward to verify that this $h_{\mu\sigma}$ is traceless, satisfies $h_{v\mu}=0$, obeys the de Donder gauge conditions, and solves the linearised Einstein equations \eqref{linefe}. In demonstrating this, it is useful to define a momentum associated with the graviton:
\begin{multline}\label{momentum}
K_{\mu}\,\d X^\mu := \d\phi_{k}=\\
 k_0\,\d v
 +\left( \frac{k_0}{2}\,\dot{\sigma}_{bc}\,x^{b}x^{c}+k_{i}\dot{E}^{i}_{b}x^{b}+\frac{k_{i}k_{j}}{2k_0}\gamma^{ij}\right)\d u+(k_{i}E^{i}_{a}+k_{0}\,\sigma_{ab}x^{b})\d x^a\,.
\end{multline}
This is null with respect to the plane wave metric ($K^2=g^{\mu\nu}K_{\mu}K_{\nu}=0$), and is also compatible with the polarization vector \eqref{grpol} in the sense that $g^{\mu\nu}\vepsilon_{\mu} K_{\nu}=0$.

\medskip

Having constructed fixed graviton vertex operators for the type II model on a plane wave background, one can now ask for vertex operators in other pictures. For the calculation of worldsheet correlation functions, it is particularly important to have the vertex operators of zero picture number (i.e., without any $\delta(\gamma)$ or $\delta(\tilde{\gamma})$ insertions), usually obtained via the fermionic descent procedure from fixed vertex operators (c.f., \cite{Friedan:1985ge}). This procedure entails successively extracting the simple poles between the fixed vertex operator and the currents $\cG$, $\tilde{\cG}$, and results in the picture number zero vertex operator, defined up to gauge transformations. 

In flat space, the pure gauge contributions are proportional to $k\cdot P$ (i.e., the scattering equations), but on a curved background they can be much more subtle. To isolate the gauge-invariant portion of the descended vertex operator, we can exploit the fact that $\{\cG,\tilde{\cG}\}=\cH$ quantum mechanically. First compute the descended operator by isolating the simple poles of $\tilde{\cG}(\cG V)$, and then again by isolating the simple poles of $\cG(\tilde{\cG}V)$. Since $\{\cG,\tilde{\cG}\}=\cH$, it follows that the sum of the two resulting operators must be pure gauge, while the difference will be the gauge-invariant contribution to the descended operator.

To do this, first compute
\be\label{gsp1}
\cG(z)\,\tilde{\psi}_{\mu}\psi^{\sigma}h^{\mu}_{\sigma}(w) \sim \frac{1}{z-w}\left(\Pi_{\mu}\,\psi^{\sigma}\,h^{\mu}_{\sigma} - \tilde{\psi}_{\mu}\psi^{\sigma}\psi^{\rho}\,\partial_{\rho}h^{\mu}_{\sigma}\right)+\cdots\,,
\ee
\begin{multline}\label{gsp2}
\tilde{\cG}(z)\,\tilde{\psi}_{\mu}\psi^{\sigma}h^{\mu}_{\sigma}(w)\sim \frac{1}{z-w}\left(-g^{\sigma\rho}\,\Pi_{\rho}\,\tilde{\psi}_{\mu}\,h^{\mu}_{\sigma} - g^{\lambda\rho}\,\tilde{\psi}_{\lambda}\tilde{\psi}_{\mu}\psi^{\sigma}\,\partial_{\rho}h^{\mu}_{\sigma}\right. \\
\left.-2 H^{a}_{b}\,x^{b}\,\tilde{\psi}_{v}\tilde{\psi}_{c}\psi^{u}\,h^{c}_{a}\right)+\cdots\,,
\end{multline}
where the $+\cdots$ indicate higher-order poles. Then take the simple pole of $\tilde{\cG}$ with \eqref{gsp1} and of $\cG$ with \eqref{gsp2}, and compute the difference. Thanks to the current algebra $\{\cG,\tilde{\cG}\}=\cH$ (which holds at the quantum level since the plane wave metric solves the vacuum equations of motion), the result is gauge invariant and BRST-closed.

A straightforward, if somewhat tedious, calculation then reveals the form of the picture number zero vertex operator:
\begin{multline}\label{dgvo}
c\,\tilde{c}\,U=c\,\tilde{c}\left[ \Pi_{\mu}\,\Pi_{\sigma}\,h^{\mu\sigma}-\Pi_{\mu}\,\tilde{\psi}_{\rho}\psi^{\sigma}\,\partial^{\rho}h^{\mu}_{\sigma} +\Pi_{\sigma}\,\tilde{\psi}_{\mu}\psi^{\rho}\,\partial_{\rho}h^{\mu\sigma} +\tilde{\psi}_{\rho}\tilde{\psi}_{\mu}\psi^{\sigma}\psi^{\lambda}\, \partial^{\rho}\partial_{\lambda} h^{\mu}_{\sigma}\right. \\
+H_{ab}\,x^{b}\,\Pi_{v}\,\tilde{\psi}_{v}\psi^{\sigma}\,h^{a}_{\sigma} + H^{a}_{b}\,x^{b}\,\Pi_{v}\,\tilde{\psi}_{c}\psi^{u}\,h^{c}_{a}-H^{a}_{b}\,x^{b}\,\Pi_{v}\,\tilde{\psi}_{v}\psi^{u}\,h^{v}_{a}-2H^{a}_{b}\,x^{b}\,\Pi_{c}\,\tilde{\psi}_{v}\psi^{c}\,h^{c}_{a} \\
+H^{a}_{b}\,\tilde{\psi}_{v}\psi^{u}\left(x^{b}\,\tilde{\psi}_{c}\psi^{d}\,\partial_{a}h^{c}_{d}-\tilde{\psi}_{c}\psi^{b}\,h^{c}_{a}-\tilde{\psi}_{a}\psi^{c}\,h^{b}_{c}-2x^{b}\,\tilde{\psi}_{c}\psi^{\rho}\,\partial_{\rho}h^{c}_{a}\right) \\
+H^{a}_{b}\,x^{b}\,\tilde{\psi}_{a}\tilde{\psi}_{\mu}\psi^{u}\psi^{c}\,\partial_{v}h^{\mu}_{c}-\frac{\dot{H}}{2}\,\tilde{\psi}_{v}\tilde{\psi}_{a}\psi^{u}\psi^{\sigma}\,\partial_{v}h^{a}_{\sigma}-H_{bc}\,x^{c}\,\tilde{\psi}_{v}\tilde{\psi}_{a}\psi^{b}\psi^{\sigma}\,\partial_{v}h^{a}_{\sigma} \\
\left.\frac{\partial H}{2}\,\tilde{\psi}_{\mu}\psi^{\sigma}\,\partial^{2}_{v}h^{\mu}_{\sigma}-\partial(H_{ab}\,x^{b}\,\tilde{\psi}_{v})\,\psi^{\sigma}\,\partial_{v}h^{a}_{\sigma}-\partial(H^{a}_{b}\,\tilde{\psi}_{v}\psi^{u})\,h^{b}_{a} + \tilde{\psi}_{\mu}\,\partial(H^{a}_{b}\,x^{b}\,\psi^{u})\,\partial_{v}h^{\mu}_{a}\right]\,.
\end{multline}
Although we have succeeded in giving an explicit formula for the descended vertex operator on a plane wave background (something which, so far, has been impossible in ordinary string theories), the result is a rather unwieldy expression. Indeed, one might worry that \eqref{dgvo} is so complicated that it is impossible to actually obtain tractable formulae for worldsheet correlation functions -- even at 3-points. Thankfully, this is not the case: many of the terms appearing in \eqref{dgvo} do not contribute to the worldsheet correlator at 3-points, and one is left with a much more manageable operator to deal with.


\subsection{3-point function}

Using the vertex operators and the explicit representation for the graviton $h_{\mu\sigma}$, we want to compute the 3-point worldsheet correlation function:
\be\label{II3p1}
\left\la V_{1}(z_1)\,V_{2}(z_2)\,c(z_3)\tilde{c}(z_3)\,U_{3}(z_3)\right\ra\,,
\ee
at genus zero, $\Sigma\cong\CP^1$. From \eqref{FixedVOBrinkmann} and the gauge condition $h_{v\mu}=0$, it follows that the fixed vertex operator $V_i$ does not contain any insertions of $\tilde{\psi}_u$ or $\psi^{v}$. This means that any insertions of $\tilde{\psi}_v$ or $\psi^u$ appearing in $U_3$ have no conjugate fields with which to Wick contract in the correlator \eqref{II3p1} due to normal ordering. Since $\tilde{\psi}_{v}(z_3)$, $\psi^{u}(z_3)$ have no zero modes at genus zero, it follows that all terms in $U_3$ which are proportional to $\tilde{\psi}_v$ or $\psi^u$ cannot contribute to the 3-point correlator.

This drastically reduces the number of terms which need to be considered in the descended vertex operator \eqref{dgvo}:
\begin{multline}\label{egU1}
U\rightarrow \Pi_{\mu}\,\Pi_{\sigma}\,h^{\mu\sigma}-\Pi_{\mu}\,\tilde{\psi}_{\rho}\psi^{\sigma}\,\partial^{\rho}h^{\mu}_{\sigma} +\Pi_{\sigma}\,\tilde{\psi}_{\mu}\psi^{\rho}\,\partial_{\rho}h^{\mu\sigma} +\tilde{\psi}_{\rho}\tilde{\psi}_{\mu}\psi^{\sigma}\psi^{\lambda}\, \partial^{\rho}\partial_{\lambda} h^{\mu}_{\sigma} \\
+\frac{\partial H}{2}\,\tilde{\psi}_{a}\psi^{b}\,\partial^{2}_{v}h^{a}_{b}\,.
\end{multline}
In fact, the term in the second line, proportional to $\partial H$, can also be discarded. Since there are no $\Pi_{\mu}$ insertions in $V_1$ or $V_2$ which can contract with $\partial H$, it follows that $H(X)$ can be treated as a function of the zero-modes of the worldsheet field $X^{\mu}$. These zero-modes are constants on the worldsheet, and thus $\partial H=0$. So for the 3-point function \eqref{II3p1}, one is able to consider an `effective' descended vertex operator:
\be\label{egU2}
U^{\mathrm{eff}}=\Pi_{\mu}\,\Pi_{\sigma}\,h^{\mu\sigma}-\Pi_{\mu}\,\tilde{\psi}_{\rho}\psi^{\sigma}\,\partial^{\rho}h^{\mu}_{\sigma} +\Pi_{\sigma}\,\tilde{\psi}_{\mu}\psi^{\rho}\,\partial_{\rho}h^{\mu\sigma} +\tilde{\psi}_{\rho}\tilde{\psi}_{\mu}\psi^{\sigma}\psi^{\lambda}\, \partial^{\rho}\partial_{\lambda} h^{\mu}_{\sigma}\,,
\ee
with $h_{\mu\sigma}$ given by \eqref{graviton}.

The ghost sector of the correlation function decouples from the matter systems, so it is easy to see that the worldsheet correlator reduces to:
\be\label{II3p2}
\left\la V_{1}(z_1)\,V_{2}(z_2)\,c(z_3)\tilde{c}(z_3)\,U_{3}(z_3)\right\ra = \frac{z_{23}^2\,z_{31}^2}{\d z_{1}\,\d z_{2}\,\d z_{3}^2}\,\left\la \tilde{\psi}_{\mu}\psi^{\sigma} h_{1\,\sigma}^{\mu}(z_1)\,\tilde{\psi}_{\rho}\psi^{\lambda} h_{2\,\lambda}^{\rho}(z_2)\,U_3^{\mathrm{eff}}(z_3)\right\ra_{\Pi X}^{\tilde{\psi}\psi}\,,
\ee
where $z_{ij}\equiv z_{i}-z_{j}$ and $\la \cdots \ra_{\Pi X}^{\tilde{\psi}\psi}$ indicates a correlation function with respect to the $(\Pi,X)$ and $(\tilde{\psi},\psi)$ worldsheet systems. To evaluate the remaining correlation function, it is useful to have explicit expressions for the (effective) vertex operators:
\be\label{IIfe}
\tilde{\psi}_{\mu}\,\psi^{\sigma}\, h_{\sigma}^{\mu}=\left(\vepsilon^{\mu}\tilde{\psi}_{\mu}\,\vepsilon_{\sigma}\psi^{\sigma}-\frac{\im}{k_0}\,\tilde{\psi}_{v}\,\psi^{u}\,\epsilon^{a}\epsilon^{b}\sigma_{ab}\right)\Omega\,\e^{\im\phi_k}\,,
\ee
and 
\begin{multline}\label{IIde}
 U^{\mathrm{eff}}=\left[(\vepsilon^{\mu}\Pi_{\mu})^2 -\frac{\im}{k_0}\Pi_v^2\,\tilde{\psi}_{v}\psi^{u}\,\epsilon^{a}\epsilon^{b}-\im\vepsilon^{\sigma}\Pi_{\sigma}\,(\tilde{\psi}_{a}\epsilon^{a}\,K_{\rho}\psi^{\rho}+K^{\mu}\tilde{\psi}_{\mu}\,\epsilon_{a}\psi^{a}) \right. \\
 -K^{\rho}\tilde{\psi}_{\rho}\,\vepsilon^{\mu}\tilde{\psi}_{\mu}\,\vepsilon_{\sigma}\psi^{\sigma}\,K_{\lambda}\psi^{\lambda} +\im k_{0}\,\tilde{\psi}_{b}\psi^{a}\sigma_{a}^{b}\,\vepsilon^{\mu}\tilde{\psi}_{\mu}\,\vepsilon_{\sigma}\psi^{\sigma} \\
 \left.+\Pi_{v}\,\epsilon^{b}\sigma_{ab}\,\vepsilon^{\mu}\tilde{\psi}_{\mu}\,\psi^{a} -\Pi_{v}\,\epsilon^{b}\sigma_{b}^{a}\,\tilde{\psi}_{a}\,\vepsilon_{\sigma}\psi^{\sigma}\right]\,\Omega\,\e^{\im\phi_k}\,,
\end{multline}
where the curved-space polarization $\vepsilon_\mu$ and momentum $K_\mu$ are as given in \eqref{grpol} and \eqref{momentum}, respectively.

Observe that in the remaining correlation function, the only $v$-dependence is in the exponentials $\e^{\im\phi_k}$ and takes the form $\exp(\im\sum_{r=1}^{3}k_{r\,0}v)$. This can be absorbed into the exponential of the action in the remaining path integral:
\be\label{react}
S_{(\Pi,X)}\rightarrow \frac{1}{2\pi}\int_{\Sigma}\Pi_{\mu}\,\dbar X^{\mu} + 2\pi\im \sum_{r=1}^{3} k_{r\,0}\,v(z)\,\delta^{2}(z-z_r)\,.
\ee
The path integral over $v$ can now be done explicitly; the zero mode integral results in a momentum conserving delta function $\delta(\sum_{r=1}^3 k_{r\,0})$, while the integral over non-zero-modes imposes an equation of motion on the conjugate field $\Pi_v$:
\be\label{pieom}
\dbar \Pi_{v}(z)=2\pi\im\sum_{r=1}^{3}k_{r\,0}\,\delta^{2}(z-z_r) \quad \Rightarrow \quad \Pi_{v}(z)=\d z\,\sum_{r=1}^{3}\frac{k_{r\,0}}{z-z_r}\,.
\ee
This allows us to replace every insertion of $\Pi_v$ in the correlator with the solution obtained from \eqref{pieom}. Similarly, there are no insertions of $\Pi_{u}$ anywhere in the remaining correlator; performing the path integral over the non-zero-modes of $u$ imposes $\dbar u=0$, reducing the worldsheet field $u(z)$ to its constant zero mode everywhere.

This leaves $(\Pi_{a}, x^a)$ as the only fields of the $(\Pi,X)$ system with non-zero-modes still in play. The only $\Pi_a$ insertions appear in $U_{3}^{\mathrm{eff}}$, while there is $x^a$-dependence lurking in the polarization components $\vepsilon_{u}$ or $\vepsilon^{v}$ as well as in the exponential factors. In the latter case, conservation of the $k_0$ momentum components (and the assumption that all three states are incoming) reduces the exponential dependence on $x^a$ to:
\be\label{expx}
\exp\left(\im E^{i}_{a}\,x^{a}\,\sum_{r=1}^{3}k_{r\,i}\right)\,.
\ee
At this point, the path integral over $\Pi_{a}$ can be done explicitly by taking all possible Wick contractions. After also taking all possible contractions in the $(\tilde{\psi},\psi)$ system, the result is the following rather unwieldy-looking expression (stripped of overall factors):
\begin{multline}\label{II3p3}
\left( \frac{\vepsilon_1 \cdot \vepsilon_2}{z_{12}}\right)^2 \left( \im \sum_{r=1}^2 \frac{\vepsilon^\mu_3  {K_r}_\mu}{z_{r3}} \right)^2
-\left(  \frac{\vepsilon_1 \cdot \vepsilon_2}{z_{12}}\right)^2 \left( \im\sigma_{ab} \sum_{r=1}^2 k_{r\,0}\frac{\epsilon^a_3 \epsilon^b_3}{z_{r3}^2} \right)
\\
+2\im\left[ \left(  \frac{\vepsilon_2 \cdot \vepsilon_3}{z_{23}}\right) \left(  \frac{\vepsilon_1 \cdot \vepsilon_2}{z_{12}}\right) \left(  \frac{\vepsilon_1\cdot {K_3}}{z_{13}}\right)  \left( \im \sum_{r=1}^2 \frac{\vepsilon_3\cdot  K_r}{z_{r3}} \right) + (1\leftrightarrow 2)\right]
\\
+2\im\left[ \left(k_{3\,0}  \frac{\vepsilon_2 \cdot \vepsilon_3}{z_{23}}\right) \left(  \frac{\vepsilon_1 \cdot \vepsilon_2}{z_{12}}\right)  \left(  \frac{\epsilon^b_{1} \sigma_{ab} \epsilon_3^a}{z_{13}^2}\right)   + (1\leftrightarrow 2)\right]
\\
+\left[ \left(  \frac{\vepsilon_2 \cdot \vepsilon_3}{z_{23}}\right)\left(  \frac{\vepsilon_2\cdot K_3}{z_{23}}\right) \left(  \frac{\vepsilon_1 \cdot K_3}{z_{13}}\right)  \left(  \frac{\vepsilon_1 \cdot \vepsilon_3}{z_{13}}\right) 
- \left(  \frac{\vepsilon_2 \cdot \vepsilon_3}{z_{23}}\right)^2 \left(  \frac{\vepsilon_1 \cdot K_3}{z_{13}}\right)^2  + (1\leftrightarrow 2)\right]
\\
- 2\left[\left( \im \sum_{r=1}^2 \frac{ {K_r}_v}{z_{r3}} \right)   \left(  \frac{\vepsilon_1 \cdot \vepsilon_2}{z_{12}}   \right)\left(  \frac{\vepsilon_2 \cdot \vepsilon_3}{z_{23}}   \right) \frac{\epsilon^b_1 \epsilon^a_3}{z_{13}}\, \sigma_{ab} + (1\leftrightarrow 2)\right]
\\
+ \im\,k_{3\,0}\left[ \left(  \frac{\vepsilon_2 \cdot \vepsilon_3}{z_{23}}   \right)^2 \frac{\epsilon_1^a \epsilon_1^b \sigma_{ab}}{z_{13}^2} -\left(  \frac{\vepsilon_2 \cdot \vepsilon_3}{z_{23}}   \right) \left(  \frac{\vepsilon_1 \cdot \vepsilon_3}{z_{13}} \right) \frac{\epsilon_1^a \epsilon_2^b \sigma_{ab}}{z_{13} z_{23}} 
 + (1\leftrightarrow 2)\right] 
\\
-\left(  \frac{\vepsilon_1 \cdot \vepsilon_2}{z_{12}}   \right)^2 \left( \frac{\epsilon^a_{3} \epsilon^b_{3}}{k_{3\,0}} \sigma_{ab}  \right) \left(\sum_{r=1}^2 \frac{ {K_r}_v}{z_{r3}} \right)^2+ \im\left[\frac{\epsilon^a_2 \epsilon^b_2}{k_{2\,0}} \sigma_{ab}\, \frac{{k_3}_0^2}{z_{23}^2}  \left(  \frac{\vepsilon_1 \cdot \vepsilon_3}{z_{13}}   \right)^2 + (1\leftrightarrow 2)\right] \,.
\end{multline}
This expression can be considerably simplified by performing the path integral over the remaining $x^a$ zero modes, which results in $d-2$ additional delta functions and a Jacobian factor:
\be\label{adelta}
\delta^{d-2}\!\left(\sum_{r=1}^3 k_{r\,i}\right)\,|E|\,,
\ee
where $|E|$ is the determinant of the vielbein $E_{i}^{a}$. 

On the support of these delta functions, and utilizing the identities
\begin{eqnarray}
\label{polrels}
K_{{r}}\cdot \varepsilon_{{s}}&=&\left\{\begin{array}{c c}
                                          0 & \mathrm{if}\;\; {r}={s} \\
                                          E^{i\,a}(k_{{r}\,0}\frac{k_{{s}\,i}}{k_{s\, 0}}\epsilon_{{s}\,a}-k_{{r}\,i}\epsilon_{{s}\,a})\quad & \mathrm{otherwise}
                                         \end{array}\right. \,,
\\
\label{polrels2}
\vepsilon_{{r}}\cdot\varepsilon_{{s}}&=&\left\{\begin{array}{c c}
                \quad    \qquad  \qquad  0 \qquad\quad\qquad& \quad \mathrm{if}\;\; {r}={s} \\
-\epsilon_{r} \cdot \epsilon_{s} &
\qquad  \mathrm{otherwise}
                                         \end{array}\right.\,,
\end{eqnarray}
the contribution \eqref{II3p3} can be massaged into a much more palatable form:
\be\label{II3p4}
 \frac{1}{z_{23}^{2}z_{31}^{2}}\left[\left(\vepsilon_1 \cdot \vepsilon_2\, \vepsilon_3\cdot K_2 +\vepsilon_2 \cdot \vepsilon_3\, \vepsilon_1 \cdot K_3+ \vepsilon_1 \cdot \vepsilon_3\, \vepsilon_2\,\cdot K_1 \right)^2 \\
 - \im  k_{1\,0}k_{2\,0}k_{3\,0}\,\sigma^{ab}\,\mathcal{C}_{a}\,\mathcal{C}_{b}\right]\,,
\ee
where 
\be\label{corr1}
\mathcal{C}_{a}:= \vepsilon_{2}\cdot\vepsilon_{3}\,\frac{\epsilon_{1\,a}}{k_{1\, 0}} + \vepsilon_{1}\cdot\vepsilon_{3}\, \frac{\epsilon_{2\,a}}{k_{2\, 0}}+\vepsilon_{1}\cdot\vepsilon_{2}\, \frac{\epsilon_{3\,a}}{k_{3\, 0}}\,,
\ee
encodes a `correction' to the tensor structure of the 3-point function in flat space-time.

These manipulations leave us with:
\begin{multline}\label{II3p5}
 \left\la \tilde{\psi}_{\mu}\psi^{\sigma} h_{1\,\sigma}^{\mu}(z_1)\,\tilde{\psi}_{\rho}\psi^{\lambda} h_{2\,\lambda}^{\rho}(z_2)\,U_3^{\mathrm{eff}}(z_3)\right\ra_{\Pi X}^{\tilde{\psi}\psi} = \delta^{d-1}\!\left(\sum_{r=1}^{3}k_r\right)\,\frac{\d z_{1}\,\d z_{2}\,\d z_{3}^2}{z_{23}^{2}z_{31}^{2}} \\
 \times \int \frac{\d u}{\sqrt{|E|}}\,\left[\left(\vepsilon_{1}\cdot\vepsilon_{3}\,K_{1}\cdot\vepsilon_{2}+ \mathrm{cyclic}\right)^{2} 
-\im\, k_{1\,0}k_{2\,0}k_{3\,0}\,\sigma^{ab}\mathcal{C}_{a}\mathcal{C}_{b}\right]\, \exp\left(\im F^{ij}\sum_{{s}=1}^{3}\frac{k_{{s}\,i}k_{{s}\,j}}{2k_{{s}\,0}}\right)\,,
\end{multline}
where 
\begin{equation*}
 \delta^{d-1}\!\left(\sum_{{r}=1}^{3}k_{{r}}\right):= \delta\!\left(\sum_{r=1}^{3} k_{r\,0}\right)\, \delta^{d-2}\!\left(\sum_{r=1}^{3} k_{r\,i}\right)\,,
\end{equation*}
encodes all of the delta functions resulting from zero mode integrations. Taking into account the ghost contributions from \eqref{II3p2}, it follows that all dependence on the vertex operator locations $z_i$ drops out (as required for M\"obius invariance of the worldsheet correlator on $\Sigma\cong\CP^1$), leaving
\begin{multline}\label{II3p6}
 \left\la V_{1}(z_1)\,V_{2}(z_2)\,c(z_3)\tilde{c}(z_3)\,U_{3}(z_3)\right\ra =\delta^{d-1}\!\left(\sum_{r=1}^{3}k_r\right)\,\int \frac{\d u}{\sqrt{|E|}}\,\left[\left(\vepsilon_{1}\cdot\vepsilon_{3}\,K_{1}\cdot\vepsilon_{2}+ \mathrm{cyclic}\right)^{2}\right. \\
 -\im\, k_{1\,0}k_{2\,0}k_{3\,0}\,\sigma^{ab}\mathcal{C}_{a}\mathcal{C}_{b}\bigg]\, \exp\left(\im F^{ij}\sum_{{s}=1}^{3}\frac{k_{{s}\,i}k_{{s}\,j}}{2k_{{s}\,0}}\right)\,.
\end{multline}
The right-hand side of this expression is equal to the 3-point amplitude for (incoming) gravitons on a plane wave space-time~\cite{Adamo:2017nia}.

\medskip

The result of this calculation can be succinctly summarized as
\be\label{II3p}
\left\la V_{1}(z_1)\,V_{2}(z_2)\,c(z_3)\tilde{c}(z_3)\,U_{3}(z_3)\right\ra=\cM_{3}^{\mathrm{pw}}(h_{1},h_{2},h_{3})\,,
\ee
where $\cM_{3}^{\mathrm{pw}}(h_{1},h_{2},h_{3})$ is the 3-point amplitude for graviton scattering on a sandwich plane wave metric. This demonstrates that the type II ambitwistor string encodes the correct interactions for gravity on non-trivial backgrounds in a practical way: the 3-point amplitude is obtained as a worldsheet correlation function with no reference whatsoever to a space-time action or Lagrangian. Furthermore, the correct linearised external states are automatically encoded by the ambitwistor string's BRST cohomology.


\section{Heterotic Model on a Gauge Field Plane Wave Background}
\label{gtPW}

Plane wave gauge fields are the gauge theory analogues of plane wave metrics. The class of such background gauge fields we consider here are highly symmetric solutions of the Yang-Mills equations valued in the Cartan of the gauge group~\cite{Adamo:2017nia}. As such, they are consistent backgrounds for the heterotic ambitwistor string. Like gravitational plane waves, all of the higher curvature invariants (e.g., $F^2$, $F^3$) of a plane wave gauge field vanish, which makes them candidate solutions to the gauge sector of the equations of motion of heterotic or type I string theory, although (to our knowledge) this fact has not been explored in the literature. 

Unlike conventional string theory on a gauge field background, we saw in Section \ref{Hcb} that the heterotic ambitwistor string remains a free worldsheet CFT. In this section we consider the quantization of the heterotic ambitwistor string on a plane wave gauge field background, deriving vertex operators in the fixed and descended pictures from the BRST cohomology and computing 3-point functions explicitly. Once more, we find that the 3-point correlation functions on the Riemann sphere reproduce the known results for 3-point gluon scattering on a plane wave gauge theory background.


\subsection{Plane wave gauge fields \& Scattering}

A plane wave gauge field in `Einstein-Rosen gauge' is one which manifests a symmetry algebra analogous to that of a plane wave metric in Einstein-Rosen coordinates. While this can be defined for generic gauge group, we restrict our attention to plane wave gauge fields valued in the Cartan subalgebra of the gauge group. With this restriction, a plane wave gauge field is given by 
\be\label{gER}
\sA=-\sA_{a}(u)\,\d x^{a}\,,
\ee
in the coordinates $\d s^2= 2\d u\d v -\d x_{a}\d x^{a}$ of Minkowski space. In further analogy with the gravitational case, a `Brinkmann gauge' version of the field is obtained by a gauge transformation of \eqref{gER}, namely $\sA\rightarrow \sA +\d(x^{a}\sA_{a})$, resulting in:
\be\label{gBr}
\sA=x^{a}\,\dot{\sA}_{a}\,\d u\,.
\ee
As the choice of terminology suggests, the Brinkmann gauge encodes the non-trivial portion of the field strength algebraically,
\be\label{gFS}
\mathsf{F}_{\mu\nu}\,\d X^\mu\wedge \d X^\nu=\dot{\sA}_{a}\d x^a \wedge \d u\,.
\ee
For this reason (and for others that arise later), we chose to work in this Brinkmann gauge. It is easy to see that \eqref{gFS} satisfies the (abelian) Yang-Mills equations.

The same subtleties that arose when considering the definition of the S-matrix on a plane wave metric background occur for plane wave background gauge fields. The resolution is identical: to obtain a well-defined scattering problem, one restricts to sandwich plane wave gauge fields, for which $\dot{\sA}_{a}(u)$ is non-vanishing only for $u_{1}\leq u\leq u_2$. This enables a sensible definition of in- and out-states. Furthermore, it can be shown that the evolution problem is unitary and there is no particle creation~\cite{Adamo:2017nia}. 

As in the gravitational case, momentum eigenstates in the in-region of the sandwich plane wave will not propagate to momentum eigenstates in the out-region, due to the gauge theoretical memory effect. To avoid any of the complications associated with this fact (which don't change the substance of the results, as in the gravitational case), we restrict our attention to the scattering of in-states only.


\subsection{Worldsheet model and Vertex operators}

On a plane wave gauge field background, the heterotic ambitwistor string is anomaly free, up to a conformal anomaly which can be eliminated with an appropriate choice of gauge group and is, in any case, irrelevant for our purposes at genus zero. The currents $\sG$ and $\sH$ of the heterotic ambitwistor string, given by \eqref{hetG}--\eqref{hetH}, take the form
\be\label{gBrG}
\sG=\Psi^{\mu}\,\Pi_{\mu}-\Psi^{u}\,x^{b}\,\dot{\sA}^{\sa}_{b}\,j^{\sa}\,,
\ee
\be\label{gBrH}
\sH=\eta^{\mu\sigma}\,\Pi_{\mu}\,\Pi_{\nu}-2\Pi_{v}\,x^{b}\,\dot{\sA}^{\sa}_{b}\,j^{\sa}+2\Psi^{b}\,\Psi^{u}\,\dot{\sA}^{\sa}_{b}\,j^{\sa}\,.
\ee
Note that there are no terms proportional to worldsheet derivatives in $\cH$, since the background gauge field obeys $\partial^{\mu} \sA_{\mu}=0$ in Brinkmann gauge. 

Equipped with the explicit BRST operator $Q$ and the free worldsheet OPEs, we can investigate the vertex operators in the BRST cohomology. We restrict our attention to vertex operators in the NS sector which should correspond to small perturbations of the background gauge field (i.e., gluons). Such vertex operators can appear with picture number $-1$ or zero. The former are fixed vertex operators; a natural ansatz for such a vertex operator is:
\begin{align}\label{gluonfvo}
V=c\,\tilde{c}\,\delta(\gamma)\, \Psi^\mu\, a_\mu^{\sa} \, j^{\sa}\,,
\end{align}
where $a_{\mu}$ is a function of $X^{\mu}$ alone, chosen to obey $a_{v}=0$. Note that although the background gauge field is valued in the Cartan subalgebra, this vertex operator carries generic colour charge with respect to the gauge group. 

Since $V$ has zero conformal weight, the only non-trivial constraints on the $Q$-closure of the ansatz \eqref{gluonfvo} come from double poles between the currents $\sG$, $\sH$ and the vertex operator. The relevant OPEs are easily seen to be:
\be\label{gG2pol}
\sG(z)\,\Psi^{\mu} a_{\mu}^{\sa} j^{\sa}(w)\sim -\frac{\partial^{\mu}a^{\sa}_{\mu}\,j^{\sa}}{(z-w)^2} + \frac{1}{z-w}\left(\cdots \right)\,,
\ee
and
\begin{multline}\label{gH2pol}
\sH(z)\,\Psi^{\mu} a_{\mu}^{\sa} j^{\sa}(w)\sim \frac{\Psi^{\mu}\,j^{\sa}}{(z-w)^2}\left[\partial_{\sigma}\partial^{\sigma} a_{\mu}^{\sa}+2 f^{\sa\mathsf{b}\mathsf{c}}\,x^{b}\,\dot{\sA}^{\mathsf{b}}_{b}\,\partial_{v}a^{\mathsf{c}}_{\mu} \right. \\
\left. +2 f^{\sa\mathsf{b}\mathsf{c}}\,\delta^{u}_{\mu}\,\dot{\sA}^{\mathsf{c}\,b}\,a^{\mathsf{b}}_{b}\right] + \frac{1}{z-w} \left(\cdots\right)\,.
\end{multline}
Vanishing of the double pole \eqref{gG2pol} enforces the Lorenz gauge condition $\partial^{\mu}a_{\mu}=0$, and vanishing of the double pole \eqref{gH2pol} imposes the linearized Yang-Mills equation for $a_{\mu}$ on the plane wave background in Lorenz gauge. Therefore, $QV=0$ for a fixed vertex operator of the form \eqref{gluonfvo} enforces precisely the conditions for $V$ to represent a gluon in Lorenz gauge with $a_v=0$ propagating on the plane wave gauge field background.

\medskip

One can now ask for an explicit representative of $a_{\mu}$ analogous to a momentum eigenstate in flat space. Such a wavefunction can be obtained from a gauge-covariant spin raising process acting on solutions to the charged wave equation on the plane wave background. The key ingredient is the function
\be\label{tildephi}
\tilde{\phi}_{k}=k_{0}\,v +\left(k_{a}+e\sA_{a}\right)\,x^{a} +\frac{f(u)}{2\,k_0}\,,
\ee
where $(k_{0},k_a)$ are $d-1$ constants parametrizing the momenta, $e$ is the charge of the gluon with respect to the background gauge field under the Cartan of the gauge group, and
\be\label{ffunc}
f(u):=\int^{u}\d s\,(k_{a}+e \sA_{a}(s))\,(k^{a}+e\sA^{a}(s))\,.
\ee
Note that $\sA_{a}$ is not gauge-invariant: the addition of a constant preserves the field strength. We take an in-state representation for which $\sA_{a}=0$ in the in-region of the sandwich plane wave (i.e., $u<u_1$) but $\sA_{a}\neq0$ as $u\rightarrow +\infty$ even though the field strength vanishes in the out region.

The gluon $a_{\mu}$ is then constructed from $\tilde{\phi}_k$ as~\cite{Adamo:2017nia}:
\be\label{gluon}
a^{\sa}_{\mu}=\mathsf{T}^{\sa}\,\tilde{\vepsilon}_{\mu}\,\e^{\im\,\tilde{\phi}_k}\,,
\ee
with the generator $\mathsf{T}^{\sa}$ of the gauge group encoding the colour charge, and the polarization given by
\begin{equation}
\label{glupol}
\tilde{\vepsilon}_\mu   \d X^\mu=\tilde{\epsilon}_{a} \left(\d x^a  +
\frac{1}{k_0}\,(k^{a}+e\sA^{a})\d u\right)
\end{equation}
Here, $\tilde{\epsilon}_{a}$ is a constant vector in $d-2$ dimensions, encoding the polarization information. The natural local  null momentum associated with the gluon is
\begin{align}\label{glumom}
{\sK}_{\mu}\,\d X^\mu & :=-\im\e^{-\im\tilde \phi_k}\,D_{\mu}\,\e^{\im\tilde{\phi}_{k}}\,\d X^\mu \nonumber \\
& = k_0\, \d v+ \frac{1}{2\,k_{0}}(k_{a}+e\sA_{a})(k^{a}+e\sA^{a})\d u +(k_{a}+e\sA_{a})\d x^a\,.
\end{align}
It is straightforward to verify that $\eta^{\mu\sigma} \sK_{\mu}\tilde{\vepsilon}_{\sigma}=\eta^{\mu\sigma}\sK_{\mu}\sK_{\sigma}=0$.

The descended gluon vertex operator (i.e., the picture number zero version of $V$) can be obtained by isolating contributions to the simple pole between $\sG$ and $V$ or equivalently by linearizing the constraint $\sH$. The resulting vertex operator is given by:
\be\label{dgluvo}
 c\,\tilde{c}\,U=c\,\tilde{c}\,j^{\sa}\left[\Pi_{\sigma}\,a^{\sa\,\sigma}-\Psi^{\sigma}\,\Psi^{\mu}\,\partial_{\sigma} a_{\mu}^{\sa}-f^{\sa\mathsf{b}\mathsf{c}}\,x^{a}\,\dot{\sA}^{\mathsf{b}}_{a}\,\Psi^{u}\,\Psi^{\mu}\,a_{\mu}^{\mathsf{c}}\right]\,.
\ee
Unlike the descended graviton vertex operator on a plane wave metric \eqref{dgvo}, this gluon vertex operator contains only one additional term relative to its flat space counterpart. The third term, proportional to $\dot{\sA}_{a}$, ensures that the resulting operator is covariant with respect to the background gauge field.


\subsection{3-point function}

The fixed and descended vertex operators can now be used to compute the 3-point correlation function on the Riemann sphere,
\be\label{H3p1}
\left\la V_{1}(z_1)\,V_{2}(z_2)\,c(z_3)\tilde{c}(z_3)\,U_{3}(z_3)\right\ra\,,
\ee
using \eqref{gluon} for an explicit representation of the incoming gluon. In order for the colour structure to produce a non-vanishing result, the sum of charges for the vertex operators under the background gauge field must vanish: $e_1+e_2+e_3=0$. 

The ghost and current algebra portions of the correlator are easily evaluated, leaving an effective correlator:
\begin{multline}\label{H3p2}
 \left\la V_{1}(z_1)\,V_{2}(z_2)\,c(z_3)\tilde{c}(z_3)\,U_{3}(z_3)\right\ra = \tr\left(\mathsf{T}_{1}\mathsf{T}_{2}\mathsf{T}_{3}\right) \frac{z_{23}\,z_{31}}{\sqrt{\d z_{1}\,\d z_{2}}\,\d z_3} \\
 \times \left\la \Psi\cdot\tilde{\vepsilon}_{1}(z_1)\,\Psi\cdot\tilde{\vepsilon}_{2}(z_2)\left(\Pi\cdot\tilde{\vepsilon}_3-\Psi\cdot\sK_{3}\,\Psi\cdot\tilde{\vepsilon}_3\right)(z_3)\,\e^{\im(\tilde{\phi}_1+\tilde{\phi}_2+\tilde{\phi}_3)}\right\ra^{\Psi\Psi}_{\Pi X}\,.
\end{multline}
The remaining correlator does not contain any insertions of $\Pi_{u}$ (since $\tilde{\vepsilon}^{u}=0$) so all $u$-dependence is immediately reduced to zero modes. This means that there are no Wick contractions into the $u$-dependent components of momenta $\sK$ or polarizations $\tilde{\varepsilon}$, or into the $u$-dependent terms appearing in the exponential through $\tilde{\phi}_k$. Since the $u$-dependence is totally relegated to zero mode integrations, the remaining fermion correlator can be seen to have the exact same structure as the 3-point function in flat space (c.f., \cite{Mason:2013sva}). 

With this in mind, it is easy to see that the remaining correlation function is reduced to:
\be\label{H3p3}
2\im\,\tr\left(\mathsf{T}_{1}\mathsf{T}_{2}\mathsf{T}_{3}\right)\,\delta^{d-1}\!\left(\sum_{r=1}^3 k_r\right) \int \d u\,\left[\tilde{\vepsilon}_{1}\cdot\tilde{\vepsilon}_3\,\sK_{1}\cdot\tilde{\vepsilon}_2+\mathrm{cyclic}\right]\,\exp\left(\im\sum_{r=1}^3\frac{f_r(u)}{2\,k_{r\,0}}\right)\,,
\ee
with the $d-1$ delta functions emerging after performing the zero mode integrals over $v$ and $x^a$. As expected, this is precisely the 3-point amplitude for gluon scattering on a gauge field plane wave background~\cite{Adamo:2017nia}:
\be\label{H3p4}
\left\la V_{1}(z_1)\,V_{2}(z_2)\,c(z_3)\tilde{c}(z_3)\,U_{3}(z_3)\right\ra=\cA_{3}^{\mathrm{pw}}(a_1,a_2,a_3)\,.
\ee
So the heterotic ambitwistor string correctly encodes the interactions of gauge theory on a curved background, with the appropriate linear perturbations (i.e., gluons) emerging from the worldsheet BRST cohomology.


\section{Discussion}
The work of \cite{Adamo:2014wea} showed that ambitwistor strings can be consistently defined on a type II supergravity background.  This suggested that it might be possible to calculate amplitudes on such backgrounds following an extension of the flat space strategy.  We have seen here that this does indeed turn out to be the case on a plane wave background at three points.  We have similarly seen that the heterotic ambitwistor model is again consistent on an abelian gauge background and that nonabelian gluon scattering can be correctly computed on plane wave analogues of such backgrounds.  The results are all checked against the three point amplitudes on plane wave backgrounds as computed in \cite{Adamo:2017nia}.  

There are many further directions to explore. The first perhaps is to go to higher numbers of points. The computations of \cite{Adamo:2017nia} were limited to three points because the propagator would be needed at four points, and that was not available in simple enough form.  In terms of the ambitwistor string in flat space-time, the new phenomenon at four points is the appearance of integrated vertex operators. These incorporate the scattering equations.  To take the calculations in this paper to four points we will therefore need to introduce some analogue of scattering equations on a curved background.  If this is successful, they will effectively encode the propagators.

Another natural direction to consider is  other backgrounds, such as (anti-) de Sitter, or black-hole or brane space-times.  These offer different challenges, with more sophisticated global issues to be addressed already in the space-time version of the calculations. In the (anti-) de Sitter case, the background is not actually a vacuum solution with respect to the equations of motion arising in the RNS-like formulation of the ambitwistor string used here. Ostensibly, AdS backgrounds would require a manifestly supersymmetric worldsheet model, such as the pure spinor formalism, where the scalar curvature of the background (times a compact space) is supported by Ramond-Ramond flux. While there has been some progress in describing the pure spinor ambitwistor string on such backgrounds (c.f. \cite{Chandia:2015sfa,Chandia:2015xfa,Azevedo:2016zod}), there is currently no formulation which is quantum mechanically consistent as a worldsheet theory. If these issues could be resolved, then it would enable computations akin to the ones performed in this paper on (A)dS background geometries. 

A further direction is to take more seriously the fact that ambitwistor strings have target ambitwistor space.  We should therefore  construct the curved ambitwistor spaces more explicitly for the curved backgrounds under consideration.   We must then learn how to quantize ambitwistor strings for amplitude calculations in such backgrounds.  

A separate question is whether the heterotic ambitwistor model is consistent on a non-Abelian background.  Our preliminary calculations indicate that this is not simply the case.  This is perhaps because of nontrivial couplings to the $R^2$ gravity theories that those models give rise to.  It would be interesting to resolve this issue.

A key theme of \cite{Adamo:2017nia} was the extent to which the double copy relationship between gravity and Yang-Mills amplitudes, as expressed for example in colour/kinematics duality \cite{Bern:2010ue}, survives in curved space.  The answer was that this is indeed the case with suitable modifications.  However, the curved space formulation of ambitwistor strings in \cite{Adamo:2014wea} was not expressed in such a way that the double copy is apparent. Finding a version of the ambitwistor string which manifests the double copy relationship on a curved background would provide further evidence that colour/kinematics duality persists in a useful way on non-trivial backgrounds.

\acknowledgments

We would like to thank Kai R\"ohrig, David Skinner, Piotr Tourkine, Arkady Tseytlin and Pedro Vieira for useful discussions. TA, EC and LM would like to thank the Kavli Institute for Theoretical Physics for hospitality while part of this work was completed; this research was supported in part by the National Science Foundation under Grant No. NSF PHY-1125915. TA is supported by an Imperial College Junior Research Fellowship; EC and LM are supported by EPSRC grant EP/M018911/1; SN is supported by EPSRC grant EP/M50659X/1 and a Studienstiftung des deutschen Volkes scholarship.


\bibliography{planwaves}

\providecommand{\href}[2]{#2}\begingroup\raggedright\begin{thebibliography}{10}

\bibitem{AlvarezGaume:1981hn}
L.~Alvarez-Gaume, D.~Z. Freedman and S.~Mukhi, \emph{{The Background Field
  Method and the Ultraviolet Structure of the Supersymmetric Nonlinear Sigma
  Model}}, \href{http://dx.doi.org/10.1016/0003-4916(81)90006-3}{\emph{Annals
  Phys.} {\bfseries 134} (1981) 85}.

\bibitem{Braaten:1985is}
E.~Braaten, T.~L. Curtright and C.~K. Zachos, \emph{{Torsion and Geometrostasis
  in Nonlinear Sigma Models}},
  \href{http://dx.doi.org/10.1016/0550-3213(85)90053-7}{\emph{Nucl. Phys.}
  {\bfseries B260} (1985) 630}.

\bibitem{Fradkin:1985ys}
E.~S. Fradkin and A.~A. Tseytlin, \emph{{Quantum String Theory Effective
  Action}}, \href{http://dx.doi.org/10.1016/0550-3213(86)90522-5,
  10.1016/0550-3213(85)90559-0}{\emph{Nucl. Phys.} {\bfseries B261} (1985)
  1--27}.

\bibitem{Callan:1985ia}
C.~G. Callan, Jr., E.~J. Martinec, M.~J. Perry and D.~Friedan, \emph{{Strings
  in Background Fields}},
  \href{http://dx.doi.org/10.1016/0550-3213(85)90506-1}{\emph{Nucl. Phys.}
  {\bfseries B262} (1985) 593--609}.

\bibitem{Banks:1986fu}
T.~Banks, D.~Nemeschansky and A.~Sen, \emph{{Dilaton Coupling and BRST
  Quantization of Bosonic Strings}},
  \href{http://dx.doi.org/10.1016/0550-3213(86)90432-3}{\emph{Nucl. Phys.}
  {\bfseries B277} (1986) 67--86}.

\bibitem{Abouelsaood:1986gd}
A.~Abouelsaood, C.~G. Callan, Jr., C.~R. Nappi and S.~A. Yost, \emph{{Open
  Strings in Background Gauge Fields}},
  \href{http://dx.doi.org/10.1016/0550-3213(87)90164-7}{\emph{Nucl. Phys.}
  {\bfseries B280} (1987) 599--624}.

\bibitem{Amati:1988sa}
D.~Amati and C.~Klimcik, \emph{{Nonperturbative Computation of the Weyl Anomaly
  for a Class of Nontrivial Backgrounds}},
  \href{http://dx.doi.org/10.1016/0370-2693(89)91092-7}{\emph{Phys. Lett.}
  {\bfseries B219} (1989) 443--447}.

\bibitem{Horowitz:1989bv}
G.~T. Horowitz and A.~R. Steif, \emph{{Space-Time Singularities in String
  Theory}}, \href{http://dx.doi.org/10.1103/PhysRevLett.64.260}{\emph{Phys.
  Rev. Lett.} {\bfseries 64} (1990) 260}.

\bibitem{Freund:1980xh}
P.~G.~O. Freund and M.~A. Rubin, \emph{{Dynamics of Dimensional Reduction}},
  \href{http://dx.doi.org/10.1016/0370-2693(80)90590-0}{\emph{Phys. Lett.}
  {\bfseries 97B} (1980) 233--235}.

\bibitem{Schwarz:1983qr}
J.~H. Schwarz, \emph{{Covariant Field Equations of Chiral N=2 D=10
  Supergravity}},
  \href{http://dx.doi.org/10.1016/0550-3213(83)90192-X}{\emph{Nucl. Phys.}
  {\bfseries B226} (1983) 269}.

\bibitem{Blau:2001ne}
M.~Blau, J.~M. Figueroa-O'Farrill, C.~Hull and G.~Papadopoulos, \emph{{A New
  maximally supersymmetric background of IIB superstring theory}},
  \href{http://dx.doi.org/10.1088/1126-6708/2002/01/047}{\emph{JHEP} {\bfseries
  01} (2002) 047}, [\href{https://arxiv.org/abs/hep-th/0110242}{{\ttfamily
  hep-th/0110242}}].

\bibitem{Blau:2002dy}
M.~Blau, J.~M. Figueroa-O'Farrill, C.~Hull and G.~Papadopoulos, \emph{{Penrose
  limits and maximal supersymmetry}},
  \href{http://dx.doi.org/10.1088/0264-9381/19/10/101}{\emph{Class. Quant.
  Grav.} {\bfseries 19} (2002) L87--L95},
  [\href{https://arxiv.org/abs/hep-th/0201081}{{\ttfamily hep-th/0201081}}].

\bibitem{Metsaev:1998it}
R.~R. Metsaev and A.~A. Tseytlin, \emph{{Type IIB superstring action in AdS(5)
  x S**5 background}},
  \href{http://dx.doi.org/10.1016/S0550-3213(98)00570-7}{\emph{Nucl. Phys.}
  {\bfseries B533} (1998) 109--126},
  [\href{https://arxiv.org/abs/hep-th/9805028}{{\ttfamily hep-th/9805028}}].

\bibitem{Metsaev:2001bj}
R.~R. Metsaev, \emph{{Type IIB Green-Schwarz superstring in plane wave
  Ramond-Ramond background}},
  \href{http://dx.doi.org/10.1016/S0550-3213(02)00003-2}{\emph{Nucl. Phys.}
  {\bfseries B625} (2002) 70--96},
  [\href{https://arxiv.org/abs/hep-th/0112044}{{\ttfamily hep-th/0112044}}].

\bibitem{Metsaev:2002re}
R.~R. Metsaev and A.~A. Tseytlin, \emph{{Exactly solvable model of superstring
  in Ramond-Ramond plane wave background}},
  \href{http://dx.doi.org/10.1103/PhysRevD.65.126004}{\emph{Phys. Rev.}
  {\bfseries D65} (2002) 126004},
  [\href{https://arxiv.org/abs/hep-th/0202109}{{\ttfamily hep-th/0202109}}].

\bibitem{Bena:2003wd}
I.~Bena, J.~Polchinski and R.~Roiban, \emph{{Hidden symmetries of the AdS(5) x
  S**5 superstring}},
  \href{http://dx.doi.org/10.1103/PhysRevD.69.046002}{\emph{Phys. Rev.}
  {\bfseries D69} (2004) 046002},
  [\href{https://arxiv.org/abs/hep-th/0305116}{{\ttfamily hep-th/0305116}}].

\bibitem{Berkovits:2004xu}
N.~Berkovits, \emph{{Quantum consistency of the superstring in AdS(5) x S**5
  background}},
  \href{http://dx.doi.org/10.1088/1126-6708/2005/03/041}{\emph{JHEP} {\bfseries
  03} (2005) 041}, [\href{https://arxiv.org/abs/hep-th/0411170}{{\ttfamily
  hep-th/0411170}}].

\bibitem{Arutyunov:2008if}
G.~Arutyunov and S.~Frolov, \emph{{Superstrings on AdS(4) x CP**3 as a Coset
  Sigma-model}},
  \href{http://dx.doi.org/10.1088/1126-6708/2008/09/129}{\emph{JHEP} {\bfseries
  09} (2008) 129}, [\href{https://arxiv.org/abs/0806.4940}{{\ttfamily
  0806.4940}}].

\bibitem{Stefanski:2008ik}
B.~Stefanski, jr, \emph{{Green-Schwarz action for Type IIA strings on AdS(4) x
  CP**3}}, \href{http://dx.doi.org/10.1016/j.nuclphysb.2008.09.015}{\emph{Nucl.
  Phys.} {\bfseries B808} (2009) 80--87},
  [\href{https://arxiv.org/abs/0806.4948}{{\ttfamily 0806.4948}}].

\bibitem{Maldacena:1997re}
J.~M. Maldacena, \emph{{The Large N limit of superconformal field theories and
  supergravity}}, \href{http://dx.doi.org/10.1023/A:1026654312961}{\emph{Int.
  J. Theor. Phys.} {\bfseries 38} (1999) 1113--1133},
  [\href{https://arxiv.org/abs/hep-th/9711200}{{\ttfamily hep-th/9711200}}].

\bibitem{Gubser:1998bc}
S.~S. Gubser, I.~R. Klebanov and A.~M. Polyakov, \emph{{Gauge theory
  correlators from noncritical string theory}},
  \href{http://dx.doi.org/10.1016/S0370-2693(98)00377-3}{\emph{Phys. Lett.}
  {\bfseries B428} (1998) 105--114},
  [\href{https://arxiv.org/abs/hep-th/9802109}{{\ttfamily hep-th/9802109}}].

\bibitem{Witten:1998qj}
E.~Witten, \emph{{Anti-de Sitter space and holography}}, {\emph{Adv. Theor.
  Math. Phys.} {\bfseries 2} (1998) 253--291},
  [\href{https://arxiv.org/abs/hep-th/9802150}{{\ttfamily hep-th/9802150}}].

\bibitem{Berenstein:2002jq}
D.~E. Berenstein, J.~M. Maldacena and H.~S. Nastase, \emph{{Strings in flat
  space and pp waves from N=4 superYang-Mills}},
  \href{http://dx.doi.org/10.1088/1126-6708/2002/04/013}{\emph{JHEP} {\bfseries
  04} (2002) 013}, [\href{https://arxiv.org/abs/hep-th/0202021}{{\ttfamily
  hep-th/0202021}}].

\bibitem{Delduc:2013qra}
F.~Delduc, M.~Magro and B.~Vicedo, \emph{{An integrable deformation of the
  $AdS_5 x S^5$ superstring action}},
  \href{http://dx.doi.org/10.1103/PhysRevLett.112.051601}{\emph{Phys. Rev.
  Lett.} {\bfseries 112} (2014) 051601},
  [\href{https://arxiv.org/abs/1309.5850}{{\ttfamily 1309.5850}}].

\bibitem{Arutyunov:2015mqj}
G.~Arutyunov, S.~Frolov, B.~Hoare, R.~Roiban and A.~A. Tseytlin, \emph{{Scale
  invariance of the $\eta$-deformed $AdS_5\times S^5$ superstring, T-duality
  and modified type II equations}},
  \href{http://dx.doi.org/10.1016/j.nuclphysb.2015.12.012}{\emph{Nucl. Phys.}
  {\bfseries B903} (2016) 262--303},
  [\href{https://arxiv.org/abs/1511.05795}{{\ttfamily 1511.05795}}].

\bibitem{Wulff:2016tju}
L.~Wulff and A.~A. Tseytlin, \emph{{Kappa-symmetry of superstring sigma model
  and generalized 10d supergravity equations}},
  \href{http://dx.doi.org/10.1007/JHEP06(2016)174}{\emph{JHEP} {\bfseries 06}
  (2016) 174}, [\href{https://arxiv.org/abs/1605.04884}{{\ttfamily
  1605.04884}}].

\bibitem{Maldacena:2000hw}
J.~M. Maldacena and H.~Ooguri, \emph{{Strings in AdS(3) and SL(2,R) WZW model
  1.: The Spectrum}}, \href{http://dx.doi.org/10.1063/1.1377273}{\emph{J. Math.
  Phys.} {\bfseries 42} (2001) 2929--2960},
  [\href{https://arxiv.org/abs/hep-th/0001053}{{\ttfamily hep-th/0001053}}].

\bibitem{Maldacena:2000kv}
J.~M. Maldacena, H.~Ooguri and J.~Son, \emph{{Strings in AdS(3) and the SL(2,R)
  WZW model. Part 2. Euclidean black hole}},
  \href{http://dx.doi.org/10.1063/1.1377039}{\emph{J. Math. Phys.} {\bfseries
  42} (2001) 2961--2977},
  [\href{https://arxiv.org/abs/hep-th/0005183}{{\ttfamily hep-th/0005183}}].

\bibitem{Maldacena:2001km}
J.~M. Maldacena and H.~Ooguri, \emph{{Strings in AdS(3) and the SL(2,R) WZW
  model. Part 3. Correlation functions}},
  \href{http://dx.doi.org/10.1103/PhysRevD.65.106006}{\emph{Phys. Rev.}
  {\bfseries D65} (2002) 106006},
  [\href{https://arxiv.org/abs/hep-th/0111180}{{\ttfamily hep-th/0111180}}].

\bibitem{Jofre:1993hd}
O.~Jofre and C.~A. Nunez, \emph{{Strings in plane wave backgrounds revisited}},
  \href{http://dx.doi.org/10.1103/PhysRevD.50.5232}{\emph{Phys. Rev.}
  {\bfseries D50} (1994) 5232--5240},
  [\href{https://arxiv.org/abs/hep-th/9311187}{{\ttfamily hep-th/9311187}}].

\bibitem{Dolan:1999dc}
L.~Dolan and E.~Witten, \emph{{Vertex operators for AdS(3) background with
  Ramond-Ramond flux}},
  \href{http://dx.doi.org/10.1088/1126-6708/1999/11/003}{\emph{JHEP} {\bfseries
  11} (1999) 003}, [\href{https://arxiv.org/abs/hep-th/9910205}{{\ttfamily
  hep-th/9910205}}].

\bibitem{Berkovits:2000yr}
N.~Berkovits and O.~Chandia, \emph{{Superstring vertex operators in an AdS(5) x
  S**5 background}},
  \href{http://dx.doi.org/10.1016/S0550-3213(00)00697-0}{\emph{Nucl. Phys.}
  {\bfseries B596} (2001) 185--196},
  [\href{https://arxiv.org/abs/hep-th/0009168}{{\ttfamily hep-th/0009168}}].

\bibitem{Chandia:2013kja}
O.~Chandia, A.~Mikhailov and B.~C. Vallilo, \emph{{A construction of integrated
  vertex operator in the pure spinor sigma-model in $AdS_5 \times S^5$}},
  \href{http://dx.doi.org/10.1007/JHEP11(2013)124}{\emph{JHEP} {\bfseries 11}
  (2013) 124}, [\href{https://arxiv.org/abs/1306.0145}{{\ttfamily 1306.0145}}].

\bibitem{Minahan:2012fh}
J.~A. Minahan, \emph{{Holographic three-point functions for short operators}},
  \href{http://dx.doi.org/10.1007/JHEP07(2012)187}{\emph{JHEP} {\bfseries 07}
  (2012) 187}, [\href{https://arxiv.org/abs/1206.3129}{{\ttfamily 1206.3129}}].

\bibitem{Bargheer:2013faa}
T.~Bargheer, J.~A. Minahan and R.~Pereira, \emph{{Computing Three-Point
  Functions for Short Operators}},
  \href{http://dx.doi.org/10.1007/JHEP03(2014)096}{\emph{JHEP} {\bfseries 03}
  (2014) 096}, [\href{https://arxiv.org/abs/1311.7461}{{\ttfamily 1311.7461}}].

\bibitem{Minahan:2014usa}
J.~A. Minahan and R.~Pereira, \emph{{Three-point correlators from string
  amplitudes: Mixing and Regge spins}},
  \href{http://dx.doi.org/10.1007/JHEP04(2015)134}{\emph{JHEP} {\bfseries 04}
  (2015) 134}, [\href{https://arxiv.org/abs/1410.4746}{{\ttfamily 1410.4746}}].

\bibitem{Berkovits:2012ps}
N.~Berkovits and T.~Fleury, \emph{{Harmonic Superspace from the $AdS_5\times
  S^5$ Pure Spinor Formalism}},
  \href{http://dx.doi.org/10.1007/JHEP03(2013)022}{\emph{JHEP} {\bfseries 03}
  (2013) 022}, [\href{https://arxiv.org/abs/1212.3296}{{\ttfamily 1212.3296}}].

\bibitem{Azevedo:2014rva}
T.~Azevedo and N.~Berkovits, \emph{{Open-closed superstring amplitudes using
  vertex operators in $\mathrm{AdS}_5 \times \mathrm{S}^5$}},
  \href{http://dx.doi.org/10.1007/JHEP02(2015)107}{\emph{JHEP} {\bfseries 02}
  (2015) 107}, [\href{https://arxiv.org/abs/1412.5921}{{\ttfamily 1412.5921}}].

\bibitem{Constable:2002hw}
N.~R. Constable, D.~Z. Freedman, M.~Headrick, S.~Minwalla, L.~Motl,
  A.~Postnikov et~al., \emph{{PP wave string interactions from perturbative
  Yang-Mills theory}},
  \href{http://dx.doi.org/10.1088/1126-6708/2002/07/017}{\emph{JHEP} {\bfseries
  07} (2002) 017}, [\href{https://arxiv.org/abs/hep-th/0205089}{{\ttfamily
  hep-th/0205089}}].

\bibitem{Spradlin:2002ar}
M.~Spradlin and A.~Volovich, \emph{{Superstring interactions in a p p wave
  background}}, \href{http://dx.doi.org/10.1103/PhysRevD.66.086004}{\emph{Phys.
  Rev.} {\bfseries D66} (2002) 086004},
  [\href{https://arxiv.org/abs/hep-th/0204146}{{\ttfamily hep-th/0204146}}].

\bibitem{Mason:2013sva}
L.~Mason and D.~Skinner, \emph{{Ambitwistor strings and the scattering
  equations}}, \href{http://dx.doi.org/10.1007/JHEP07(2014)048}{\emph{JHEP}
  {\bfseries 07} (2014) 048},
  [\href{https://arxiv.org/abs/1311.2564}{{\ttfamily 1311.2564}}].

\bibitem{Ohmori:2015sha}
K.~Ohmori, \emph{{Worldsheet Geometries of Ambitwistor String}},
  \href{http://dx.doi.org/10.1007/JHEP06(2015)075}{\emph{JHEP} {\bfseries 06}
  (2015) 075}, [\href{https://arxiv.org/abs/1504.02675}{{\ttfamily
  1504.02675}}].

\bibitem{Casali:2015vta}
E.~Casali, Y.~Geyer, L.~Mason, R.~Monteiro and K.~A. Roehrig, \emph{{New
  Ambitwistor String Theories}},
  \href{http://dx.doi.org/10.1007/JHEP11(2015)038}{\emph{JHEP} {\bfseries 11}
  (2015) 038}, [\href{https://arxiv.org/abs/1506.08771}{{\ttfamily
  1506.08771}}].

\bibitem{Azevedo:2017lkz}
T.~Azevedo and O.~T. Engelund, \emph{{Ambitwistor formulations of $R^2$ gravity
  and $(DF)^2$ gauge theories}},
  \href{https://arxiv.org/abs/1707.02192}{{\ttfamily 1707.02192}}.

\bibitem{Cachazo:2013iea}
F.~Cachazo, S.~He and E.~Y. Yuan, \emph{{Scattering of Massless Particles:
  Scalars, Gluons and Gravitons}},
  \href{http://dx.doi.org/10.1007/JHEP07(2014)033}{\emph{JHEP} {\bfseries 07}
  (2014) 033}, [\href{https://arxiv.org/abs/1309.0885}{{\ttfamily 1309.0885}}].

\bibitem{Cachazo:2014xea}
F.~Cachazo, S.~He and E.~Y. Yuan, \emph{{Scattering Equations and Matrices:
  From Einstein To Yang-Mills, DBI and NLSM}},
  \href{http://dx.doi.org/10.1007/JHEP07(2015)149}{\emph{JHEP} {\bfseries 07}
  (2015) 149}, [\href{https://arxiv.org/abs/1412.3479}{{\ttfamily 1412.3479}}].

\bibitem{Adamo:2013tsa}
T.~Adamo, E.~Casali and D.~Skinner, \emph{{Ambitwistor strings and the
  scattering equations at one loop}},
  \href{http://dx.doi.org/10.1007/JHEP04(2014)104}{\emph{JHEP} {\bfseries 04}
  (2014) 104}, [\href{https://arxiv.org/abs/1312.3828}{{\ttfamily 1312.3828}}].

\bibitem{Casali:2014hfa}
E.~Casali and P.~Tourkine, \emph{{Infrared behaviour of the one-loop scattering
  equations and supergravity integrands}},
  \href{http://dx.doi.org/10.1007/JHEP04(2015)013}{\emph{JHEP} {\bfseries 04}
  (2015) 013}, [\href{https://arxiv.org/abs/1412.3787}{{\ttfamily 1412.3787}}].

\bibitem{Adamo:2015hoa}
T.~Adamo and E.~Casali, \emph{{Scattering equations, supergravity integrands,
  and pure spinors}},
  \href{http://dx.doi.org/10.1007/JHEP05(2015)120}{\emph{JHEP} {\bfseries 05}
  (2015) 120}, [\href{https://arxiv.org/abs/1502.06826}{{\ttfamily
  1502.06826}}].

\bibitem{Geyer:2015bja}
Y.~Geyer, L.~Mason, R.~Monteiro and P.~Tourkine, \emph{{Loop Integrands for
  Scattering Amplitudes from the Riemann Sphere}},
  \href{http://dx.doi.org/10.1103/PhysRevLett.115.121603}{\emph{Phys. Rev.
  Lett.} {\bfseries 115} (2015) 121603},
  [\href{https://arxiv.org/abs/1507.00321}{{\ttfamily 1507.00321}}].

\bibitem{Geyer:2015jch}
Y.~Geyer, L.~Mason, R.~Monteiro and P.~Tourkine, \emph{{One-loop amplitudes on
  the Riemann sphere}},
  \href{http://dx.doi.org/10.1007/JHEP03(2016)114}{\emph{JHEP} {\bfseries 03}
  (2016) 114}, [\href{https://arxiv.org/abs/1511.06315}{{\ttfamily
  1511.06315}}].

\bibitem{Geyer:2016wjx}
Y.~Geyer, L.~Mason, R.~Monteiro and P.~Tourkine, \emph{{Two-Loop Scattering
  Amplitudes from the Riemann Sphere}},
  \href{http://dx.doi.org/10.1103/PhysRevD.94.125029}{\emph{Phys. Rev.}
  {\bfseries D94} (2016) 125029},
  [\href{https://arxiv.org/abs/1607.08887}{{\ttfamily 1607.08887}}].

\bibitem{Casali:2016atr}
E.~Casali and P.~Tourkine, \emph{{On the null origin of the ambitwistor
  string}}, \href{http://dx.doi.org/10.1007/JHEP11(2016)036}{\emph{JHEP}
  {\bfseries 11} (2016) 036},
  [\href{https://arxiv.org/abs/1606.05636}{{\ttfamily 1606.05636}}].

\bibitem{Casali:2017zkz}
E.~Casali, Y.~Herfray and P.~Tourkine, \emph{{The complex null string, Galilean
  conformal algebra and scattering equations}},
  \href{https://arxiv.org/abs/1707.09900}{{\ttfamily 1707.09900}}.

\bibitem{Geyer:2014fka}
Y.~Geyer, A.~E. Lipstein and L.~J. Mason, \emph{{Ambitwistor Strings in Four
  Dimensions}},
  \href{http://dx.doi.org/10.1103/PhysRevLett.113.081602}{\emph{Phys. Rev.
  Lett.} {\bfseries 113} (2014) 081602},
  [\href{https://arxiv.org/abs/1404.6219}{{\ttfamily 1404.6219}}].

\bibitem{Bandos:2014lja}
I.~Bandos, \emph{{Twistor/ambitwistor strings and null-superstrings in
  spacetime of D=4, 10 and 11 dimensions}},
  \href{http://dx.doi.org/10.1007/JHEP09(2014)086}{\emph{JHEP} {\bfseries 09}
  (2014) 086}, [\href{https://arxiv.org/abs/1404.1299}{{\ttfamily 1404.1299}}].

\bibitem{Adamo:2014yya}
T.~Adamo, E.~Casali and D.~Skinner, \emph{{Perturbative gravity at null
  infinity}},
  \href{http://dx.doi.org/10.1088/0264-9381/31/22/225008}{\emph{Class. Quant.
  Grav.} {\bfseries 31} (2014) 225008},
  [\href{https://arxiv.org/abs/1405.5122}{{\ttfamily 1405.5122}}].

\bibitem{Geyer:2014lca}
Y.~Geyer, A.~E. Lipstein and L.~Mason, \emph{{Ambitwistor strings at null
  infinity and (subleading) soft limits}},
  \href{http://dx.doi.org/10.1088/0264-9381/32/5/055003}{\emph{Class. Quant.
  Grav.} {\bfseries 32} (2015) 055003},
  [\href{https://arxiv.org/abs/1406.1462}{{\ttfamily 1406.1462}}].

\bibitem{Lipstein:2015rxa}
A.~E. Lipstein, \emph{{Soft Theorems from Conformal Field Theory}},
  \href{http://dx.doi.org/10.1007/JHEP06(2015)166}{\emph{JHEP} {\bfseries 06}
  (2015) 166}, [\href{https://arxiv.org/abs/1504.01364}{{\ttfamily
  1504.01364}}].

\bibitem{Adamo:2015fwa}
T.~Adamo and E.~Casali, \emph{{Perturbative gauge theory at null infinity}},
  \href{http://dx.doi.org/10.1103/PhysRevD.91.125022}{\emph{Phys. Rev.}
  {\bfseries D91} (2015) 125022},
  [\href{https://arxiv.org/abs/1504.02304}{{\ttfamily 1504.02304}}].

\bibitem{Berkovits:2013xba}
N.~Berkovits, \emph{{Infinite Tension Limit of the Pure Spinor Superstring}},
  \href{http://dx.doi.org/10.1007/JHEP03(2014)017}{\emph{JHEP} {\bfseries 03}
  (2014) 017}, [\href{https://arxiv.org/abs/1311.4156}{{\ttfamily 1311.4156}}].

\bibitem{Chandia:2015sfa}
O.~Chandia and B.~C. Vallilo, \emph{{Ambitwistor pure spinor string in a type
  II supergravity background}},
  \href{http://dx.doi.org/10.1007/JHEP06(2015)206}{\emph{JHEP} {\bfseries 06}
  (2015) 206}, [\href{https://arxiv.org/abs/1505.05122}{{\ttfamily
  1505.05122}}].

\bibitem{Jusinskas:2016qjd}
R.~L. Jusinskas, \emph{{Notes on the ambitwistor pure spinor string}},
  \href{http://dx.doi.org/10.1007/JHEP05(2016)116}{\emph{JHEP} {\bfseries 05}
  (2016) 116}, [\href{https://arxiv.org/abs/1604.02915}{{\ttfamily
  1604.02915}}].

\bibitem{Reid-Edwards:2015stz}
R.~A. Reid-Edwards, \emph{{Ambitwistor String Theory in the Operator
  Formalism}}, \href{http://dx.doi.org/10.1007/JHEP06(2016)084}{\emph{JHEP}
  {\bfseries 06} (2016) 084},
  [\href{https://arxiv.org/abs/1511.08406}{{\ttfamily 1511.08406}}].

\bibitem{Reid-Edwards:2017goq}
R.~A. Reid-Edwards and D.~A. Riccombeni, \emph{{A Superstring Field Theory for
  Supergravity}},  \href{https://arxiv.org/abs/1701.05495}{{\ttfamily
  1701.05495}}.

\bibitem{Adamo:2017zkm}
T.~Adamo, R.~Monteiro and M.~F. Paulos, \emph{{Space-time CFTs from the Riemann
  sphere}},  \href{https://arxiv.org/abs/1703.04589}{{\ttfamily 1703.04589}}.

\bibitem{Adamo:2014wea}
T.~Adamo, E.~Casali and D.~Skinner, \emph{{A Worldsheet Theory for
  Supergravity}}, \href{http://dx.doi.org/10.1007/JHEP02(2015)116}{\emph{JHEP}
  {\bfseries 02} (2015) 116},
  [\href{https://arxiv.org/abs/1409.5656}{{\ttfamily 1409.5656}}].

\bibitem{Adamo:2015ina}
T.~Adamo, \emph{{Gravity with a cosmological constant from rational curves}},
  \href{http://dx.doi.org/10.1007/JHEP11(2015)098}{\emph{JHEP} {\bfseries 11}
  (2015) 098}, [\href{https://arxiv.org/abs/1508.02554}{{\ttfamily
  1508.02554}}].

\bibitem{Adamo:2017nia}
T.~Adamo, E.~Casali, L.~Mason and S.~Nekovar, \emph{{Scattering on plane waves
  and the double copy}},  \href{https://arxiv.org/abs/1706.08925}{{\ttfamily
  1706.08925}}.

\bibitem{Lebrun:1983pa}
C.~LeBrun, \emph{Spaces of complex null geodesics in complex-riemannian
  geometry}, {\emph{Transactions of the American Mathematical Society}
  {\bfseries 278} (1983) 209--231}.

\bibitem{einstein1916}
A.~{Einstein}, \emph{{N{\"a}herungsweise Integration der Feldgleichungen der
  Gravitation}}, {\emph{Sitzungsberichte der K{\"o}niglich Preu{\ss}ischen
  Akademie der Wissenschaften (Berlin), Seite 688-696.} (1916) }.

\bibitem{Baldwin:1926}
O.~R. Baldwin and G.~B. Jeffery, \emph{{The relativity theory of plane waves}},
  {\emph{Proc.Roy.Soc.Lond.} {\bfseries A111} (1926) 95}.

\bibitem{Ehlers:1962zz}
J.~Ehlers and W.~Kundt, \emph{{Exact solutions of the gravitational field
  equations}},  in \emph{{Gravitation, An Introduction to Current Research}}
  (L.~Witten, ed.), p.~49.
\newblock Wiley: New York, 1962.

\bibitem{griffiths1991colliding}
J.~Griffiths, \emph{Colliding plane waves in general relativity}.
\newblock Oxford mathematical monographs. Clarendon Press, 1991.

\bibitem{Stephani:2003tm}
H.~Stephani, D.~Kramer, M.~A.~H. MacCallum, C.~Hoenselaers and E.~Herlt,
  \emph{{Exact solutions of Einstein's field equations}}.
\newblock Cambridge University Press, 2~ed., 2004.

\bibitem{Blau:2011}
M.~Blau, \emph{{Plane waves and Penrose limits}},  tech. rep., Universit\'{e}
  de Neuch\^{a}tel, 2011.

\bibitem{Einstein:1937qu}
A.~Einstein and N.~Rosen, \emph{{On Gravitational waves}},
  \href{http://dx.doi.org/10.1016/S0016-0032(37)90583-0}{\emph{J. Franklin
  Inst.} {\bfseries 223} (1937) 43--54}.

\bibitem{Brinkmann:1925fr}
H.~W. Brinkmann, \emph{{Einstein spapces which are mapped conformally on each
  other}}, \href{http://dx.doi.org/10.1007/BF01208647}{\emph{Math. Ann.}
  {\bfseries 94} (1925) 119--145}.

\bibitem{Penrose:1965rx}
R.~Penrose, \emph{{A Remarkable property of plane waves in general
  relativity}}, \href{http://dx.doi.org/10.1103/RevModPhys.37.215}{\emph{Rev.
  Mod. Phys.} {\bfseries 37} (1965) 215--220}.

\bibitem{Bondi:1989vm}
H.~Bondi and F.~A.~E. Pirani, \emph{{Gravitational Waves in General Relativity.
  13: Caustic Property of Plane Waves}},
  \href{http://dx.doi.org/10.1098/rspa.1989.0016}{\emph{Proc. Roy. Soc. Lond.}
  {\bfseries A421} (1989) 395--410}.

\bibitem{Bondi:1958aj}
H.~Bondi, F.~A.~E. Pirani and I.~Robinson, \emph{{Gravitational waves in
  general relativity. 3. Exact plane waves}},
  \href{http://dx.doi.org/10.1098/rspa.1959.0124}{\emph{Proc. Roy. Soc. Lond.}
  {\bfseries A251} (1959) 519--533}.

\bibitem{Gibbons:1975jb}
G.~W. Gibbons, \emph{{Quantized Fields Propagating in Plane Wave Space-Times}},
  \href{http://dx.doi.org/10.1007/BF01629249}{\emph{Commun. Math. Phys.}
  {\bfseries 45} (1975) 191--202}.

\bibitem{Garriga:1990dp}
J.~Garriga and E.~Verdaguer, \emph{{Scattering of quantum particles by
  gravitational plane waves}},
  \href{http://dx.doi.org/10.1103/PhysRevD.43.391}{\emph{Phys. Rev.} {\bfseries
  D43} (1991) 391--401}.

\bibitem{Friedlander:1975eqa}
F.~G. Friedlander, \emph{{The Wave Equation on a Curved Space-Time}}.
\newblock Cambridge University Press, 1975.

\bibitem{Ward:1987ws}
R.~S. Ward, \emph{{Progressing waves in flat space-time and in plane wave
  space-times}},
  \href{http://dx.doi.org/10.1088/0264-9381/4/3/034}{\emph{Class. Quant. Grav.}
  {\bfseries 4} (1987) 775--778}.

\bibitem{Mason:1989}
L.~J. Mason, \emph{{On Ward's integral formula for the wave equation in plane
  wave space-times}}, {\emph{Twistor Newsletter} {\bfseries 28} (1989) 17--19}.

\bibitem{Friedan:1985ge}
D.~Friedan, E.~J. Martinec and S.~H. Shenker, \emph{{Conformal Invariance,
  Supersymmetry and String Theory}},
  \href{http://dx.doi.org/10.1016/0550-3213(86)90356-1,
  10.1016/S0550-3213(86)80006-2}{\emph{Nucl. Phys.} {\bfseries B271} (1986)
  93--165}.

\bibitem{Chandia:2015xfa}
O.~Chandia and B.~C. Vallilo, \emph{{On-shell type II supergravity from the
  ambitwistor pure spinor string}},
  \href{http://dx.doi.org/10.1088/0264-9381/33/18/185003}{\emph{Class. Quant.
  Grav.} {\bfseries 33} (2016) 185003},
  [\href{https://arxiv.org/abs/1511.03329}{{\ttfamily 1511.03329}}].

\bibitem{Azevedo:2016zod}
T.~Azevedo and R.~L. Jusinskas, \emph{{Background constraints in the infinite
  tension limit of the heterotic string}},
  \href{http://dx.doi.org/10.1007/JHEP08(2016)133}{\emph{JHEP} {\bfseries 08}
  (2016) 133}, [\href{https://arxiv.org/abs/1607.06805}{{\ttfamily
  1607.06805}}].

\bibitem{Bern:2010ue}
Z.~Bern, J.~J.~M. Carrasco and H.~Johansson, \emph{{Perturbative Quantum
  Gravity as a Double Copy of Gauge Theory}},
  \href{http://dx.doi.org/10.1103/PhysRevLett.105.061602}{\emph{Phys. Rev.
  Lett.} {\bfseries 105} (2010) 061602},
  [\href{https://arxiv.org/abs/1004.0476}{{\ttfamily 1004.0476}}].

\end{thebibliography}\endgroup
\bibliographystyle{JHEP}

\end{document}